\documentclass[utf8,twocolumn]{article}
\usepackage{framed,multirow}

\usepackage{amssymb}
\usepackage{latexsym}

\usepackage{url}
\usepackage{xcolor}
\usepackage{fullpage}
\usepackage{authblk}

\usepackage{hyperref}
\usepackage{amsmath,amssymb,amsfonts}
\usepackage{graphicx}
\usepackage{textcomp}
\usepackage{tabularx}
\usepackage[ruled]{algorithm2e}
\usepackage{booktabs}

\usepackage[symbols,nogroupskip,nonumberlist,sort=use]{glossaries-extra}

\definecolor{newcolor}{rgb}{.8,.349,.1}

\makenoidxglossaries

\glsxtrnewsymbol[description={
Thresholds
}]{thrs}{\ensuremath{T}}

\glsxtrnewsymbol[description={
Fundus image
}]{image}{\ensuremath{I}}

\glsxtrnewsymbol[description={
Vessel segmentation ground truth
}]{vessel-gt}{\ensuremath{I_p}}

\glsxtrnewsymbol[description={
Vessel segmentation Likelihood(probability map)
}]{vessel-pmap}{\ensuremath{\hat{I}_{p}}}

\glsxtrnewsymbol[description={
AV ground truth
}]{av-gt}{\ensuremath{I_{av}}}

\glsxtrnewsymbol[description={
AV likelihood(prior)
}]{av-pmap}{\ensuremath{\hat{I}_{av}}}

\glsxtrnewsymbol[description={
Min diameter for connected comp to be a part of vessels.
}]{diam_lim}{\ensuremath{d}}

\glsxtrnewsymbol[description={
Weights by vessel pmap in Dijks
}]{w_vessel}{\ensuremath{w_v}}

\glsxtrnewsymbol[description={
Weights by pixel color in Lab space for Dijks
}]{w_color}{\ensuremath{w_c}}

\glsxtrnewsymbol[description={
Weights by pixel distance from background in Dijks
}]{w_width}{\ensuremath{w_w}}

\glsxtrnewsymbol[description={
A vessel path repel factor. Controls how closely two vessel can be traced such two near passing vessels are traced without touching.
}]{dist_r}{\ensuremath{\Gamma}}

\glsxtrnewsymbol[description={
Min number of nodes for a segment.
}]{seg_len_lim}{\ensuremath{L}}

\glsxtrnewsymbol[description={
Number of iterations of re Dijsk
}]{dijks2_iters}{\ensuremath{D_I}}

\glsxtrnewsymbol[description={
Dijks k-expansion.
}]{dijks2_k_nbrs}{\ensuremath{k}}

\glsxtrnewsymbol[description={
False seg lim
}]{false_seg_len}{\ensuremath{l}}

\glsxtrnewsymbol[description={
Cost jump for prop
}]{c_jump}{\ensuremath{\Delta}}

\glsxtrnewsymbol[description={
ONH radius
}]{onh_r}{\ensuremath{r}}

\glsxtrnewsymbol[description={
Default Branch forward pair theta coefficient
}]{b_th}{\ensuremath{b_{th}}}

\glsxtrnewsymbol[description={
Default Branch forward pair theta coefficient for A/V Pair
}]{a_th}{\ensuremath{a_{th}}}

\glsxtrnewsymbol[description={
Labeled Branch-forward segments pair A/V cross-entropy coefficient
}]{a_ce}{\ensuremath{a_{ce}}}

\glsxtrnewsymbol[description={
$CR_{pair}$ Branch cost scale factor.
}]{a_1}{\ensuremath{a_1}}

\glsxtrnewsymbol[description={
$CR_{Prob}$ Branch cost scale factor.
}]{a_2}{\ensuremath{a_2}}

\glsxtrnewsymbol[description={
Branchness cost (BR) scale factor.
}]{a_3}{\ensuremath{a_3}}

\glsxtrnewsymbol[description={
Labeled branch-forward graph cost scale factor.
}]{a_4}{\ensuremath{a_4}}

\glsxtrnewsymbol[description={
Epsilon
}]{eps}{\ensuremath{\epsilon}}

\glsxtrnewsymbol[description={
Prop threshold for AV
}]{prop_score_thr}{\ensuremath{t_{av}}}

\begin{document}

% \verso{Aashis Khanal \textit{et~al.}}

% \begin{frontmatter}

\title{Fully automated tree topology estimation and artery-vein classification}

\date{}

% \author[1]{Aashis \snm{Khanal}}
% \cortext[cor1]{Corresponding author: 
%   Tel.: +0-000-000-0000;  
%   fax: +0-000-000-0000;}
% \author[2]{Saeid \snm{Motevali}}
% \author[3]{Rolando \snm{Estrada}\corref{cor1}}

% \address[1]{Georgia State University,  Atlanta, GA 30303 USA}

% \received{XX XXX XXXX}
% \finalform{XX XXX XXXX}
% \accepted{XX XXX XXXX}
% \availableonline{XX XXX XXXX}
% \communicated{XXX}

\author{Aashis Khanal}
\author{Saeid Motevali}
\author{Rolando Estrada}
\affil{\texttt{\{akhanal1, smotevalialamoti1\}@student.gsu.edu, restrada1@gsu.edu}}

\affil{Department of Computer Science, Georgia State University, GA, USA}

\maketitle

\begin{abstract}

We present a fully automatic, graph-based technique for extracting the retinal vascular topology---that is, how different vessels are connected to each other---given a single color fundus image. Determining this connectivity is very challenging because vessels cross each other in a 2D image, obscuring their true paths. We quantitatively validated the usefulness of our extraction method by using it to achieve comparable state-of-the-art results in retinal artery-vein classification. Our proposed approach works as follows: We first segment the retinal vessels using our previously developed state-of-the-art segmentation method. Then, we estimate an initial graph from the extracted vessels and assign the most likely blood flow to each edge. We then use a handful of high-level operations (HLOs) to fix errors in the graph. These HLOs include detaching neighboring nodes, shifting the endpoints of an edge, and reversing the estimated blood flow direction for a branch. We use a novel cost function to find the optimal set of HLO operations for a given graph. Finally, we show that our extracted vascular structure is correct by propagating artery/vein labels along the branches. As our experiments show, our topology-based artery-vein labeling achieved state-of-the-art results on three datasets: DRIVE, AV-WIDE, and INSPIRE. We also performed several ablation studies to separately verify the importance of the segmentation and AV labeling steps of our proposed method. These ablation studies further confirmed that our graph extraction pipeline correctly models the underlying vascular anatomy.
\end{abstract}

% \begin{keyword}
% %% MSC codes here, in the form: \MSC code \sep code
% %% or \MSC[2008] code \sep code (2000 is the default)
% % \MSC 41A05\sep 41A10\sep 65D05\sep 65D17
% %% Keywords
% \KWD Retinal vessel analysis\sep Deep learning\sep Graph estimation
% \end{keyword}

% \end{frontmatter}

\section{Introduction}
\label{sec:introduction}
Retinal fundus images allow ophthalmologists to diagnose a variety of ocular and cardiovascular diseases, including diabetic retinopathy (DR) \cite{retinal_imaging_and_analysis-5660089}, glaucoma \cite{economic_analysis_pmid10868871}, age-related macular degeneration (AMD) \cite{AMD_severity_scale-Ferris2005}, and the likelihood of stroke \cite{retinal_signs_and_stroke_baker_retinal_2008, Kipli2018}. These diagnoses are based on tell-tale features in the fundus image that are correlated with a higher likelihood of certain diseases. For example, a higher cup-to-disk ratio in the optic nerve is correlated with a higher likelihood of glaucoma \cite{cup-to-disc-glaucoma-BOCK2010471}, while the artery-vein ratio can be used to predict a patient's risk for diabetic retinopathy and hypertension \cite{retinal_imaging_and_analysis-5660089, hyper_pakter2001detection, av_ratio_10.1016/j.amjhyper.2004.10.011, retina_vascular_calliber_ikram_2013}. Other diagnostically useful features include vessel tortuosity, bifurcation, branching angles, and the presence of exudates \cite{10.1093/eurheartj/ehm221, 10.1167/iovs.03-1390, doi:10.1117/12.708469, Lue31, retinal_imaging_and_analysis-5660089, exutates_DR-JOSHI20181454, turtuosity_use_HART1999239}. 

\begin{figure*}[ht!]
    \centering
    \includegraphics[width=0.99\textwidth]{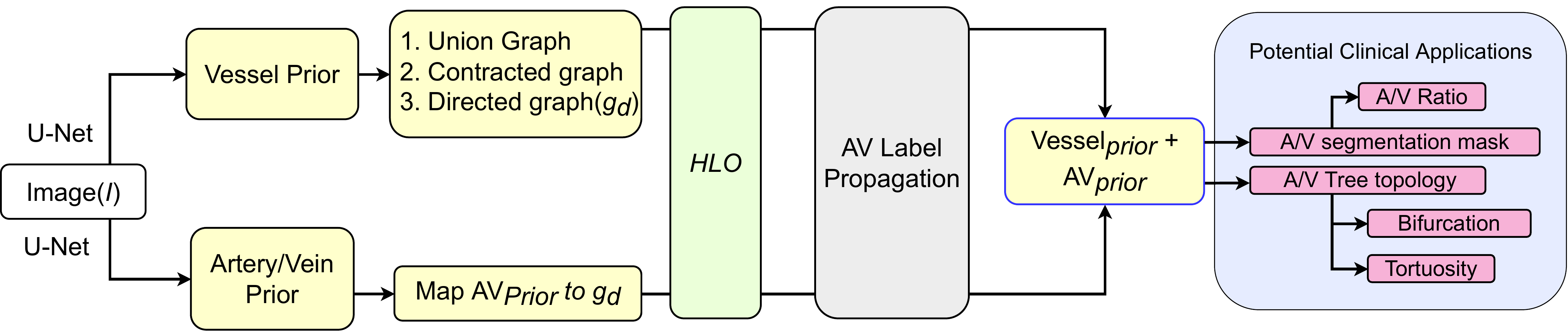}
    \caption{Pipeline flowchart: We start from a single color fundus image, then generate a vessel probability map (Vessel prior, Fig.~\ref{fig-ImagesPrior}(c)) and an artery-vein probability map (AV prior, Fig.~\ref{fig-ImagesPrior}(e)) from two separate U-Net CNNs. We then use multilevel skeletonization (Alg.~\ref{algo1:union-graph}) to produce a union graph of the vasculature (Fig.~\ref{fig-directed} (a \& b)) which is further pruned significantly using a graph contraction technique (Alg.~\ref{algo2:union-graph-contraction} and Fig.~\ref{fig-directed}(c \& d)). Afterwards, we use Dijkstra's shortest path algorithm to extract an undirected topology graph, which is assigned edge directions by our flow assignment algorithm to yield a directed graph($g_d$). We then map the $AV_{prior}$ labels to $g_d$ as shown in Fig.~\ref{fig-22-test-Mappings}(a) and perform a series of high level graph operations (HLOs) to minimize a topology cost function. Finally, we perform a simple AV label propagation step along the estimated branches. As our experiments show, our estimated topology allows us to achieve better AV classification results than a deep network alone, opening the door for a number of potential clinical applications. %\textbf{Ablation}: We perform heavy ablation studies(shown by blocks in light blue color) by replacing vessel prior with manual vessel segmentation, and AV prior by AV manual segmentation, alternating one another such that at a given point we have at least one prior in the pipeline. The pipeline works best when we have good priors(Vessel segmentation and AV label priors), as such we can always improve our priors and come back to obtain better performance.
    }
    \label{fig-flowchart}. 
\end{figure*}

Vascular features, in particular, require analyzing the properties of individual vessels, which in turn requires inferring the topology or connectivity of the underlying vasculature from the image \cite{graph_concepts_PATTON200699, retina_vascular_calliber_ikram_2013}. For example, in order to calculate the artery-vein ratio, we need to (1) identify individual vessels, (2) classify each individual vessel as either artery or vein, and (3) measure the width of the six largest arteries and veins in the region of interest. Most automated methods, however, only provide a binary segmentation of the retinal vessels, i.e., whether a given pixel is part of a vessel or not. While some topology extraction methods exist, they either require significant manual input \cite{wide_dataset:6987362} or are limited to only the main vessels in the image \cite{Amil2019,8754802}. The second limitation is of particular concern for early screening since many retinal and cardiovascular diseases affect smaller vessels earlier than larger ones.

% limitation renders the latter methods unsuitable for early screening, since Detecting the topology of small vessels is important for early screening since 

% , which by itself is not enough to estimate the aforementioned vascular features

% ; they do not estimate any topology at all.

In this paper, we present a fully automatic vessel topology extraction method that combines state-of-the-art deep learning with domain-specific graph editing techniques. In contrast to existing topology extraction methods (see Sec.~\ref{sec:av-prior-extraction}), our approach can extract the topology of the entire vasculature---not just the main vessels---without any manual input. It also estimates artery-vein labels for all vessels in the image. As we show in Sec.~\ref{sec:experiments-and-results}, our topology-based approach achieves state-of-the-art artery-vein classification results on multiple datasets.

\begin{figure*}[ht!]
    \centering
    \includegraphics[width=0.99\textwidth]{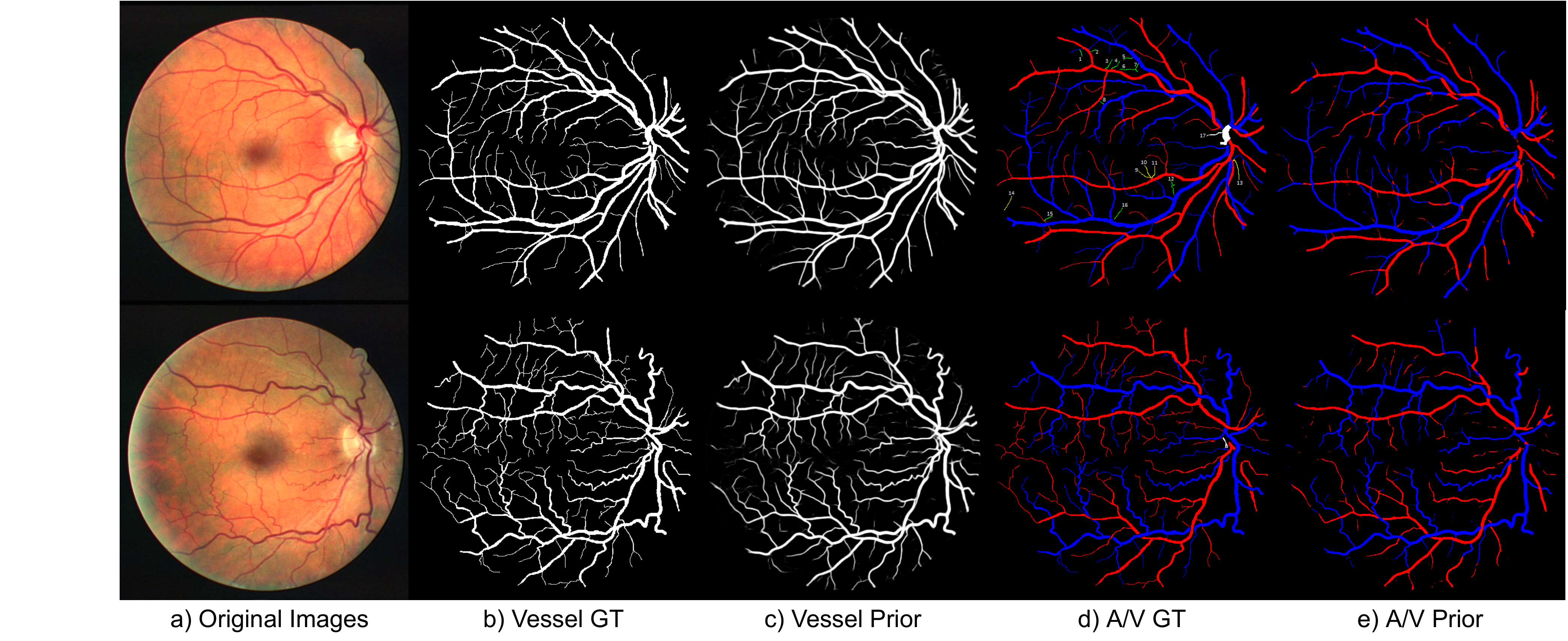}
    \caption{Deep learning priors: We use U-net-style networks to obtain an initial set of vessel and artery-vein segmentation maps. Above are the results for two sample images from the DRIVE dataset \cite{drive_dataset}. (a) Original images (b) Ground-truth vessel segmentations. (c) Vessel priors obtained from one U-Net network (d) AV ground-truth priors. (e) AV priors generated using a second U-Net network.}
    \label{fig-ImagesPrior}
\end{figure*}

Figure~\ref{fig-flowchart} shows a flowchart of our proposed method. In short, we first use two U-net-style  architectures \cite{Ronneberger2015UNetCN} to obtain (1) a pixel-level binary segmentation and (2) an initial set of artery-vein labels for the pixels identified as vascular (see Fig.~\ref{fig-ImagesPrior} for examples). We then estimate a graph from the binary segmentation, as follows. First, we use a novel technique called \textit{multilevel-skeletonization} that uses a set of thresholds ranging from low-to high to capture vessels at different scales (see Sec.~\ref{sec:multi-skel-unionG} for details). We convert this combined skeleton into an initial \textit{union graph}. This union graph captures the general shape of the vasculature well, but it contains many spurious branch nodes, as illustrated in Fig.~\ref{fig-directed}(a). To correct this, we use a novel pruning technique called union-graph-contraction that removes these spurious node and edges while keeping the overall vascular structure intact (see Fig.~\ref{fig-directed}(c) for an example). We then run Dijkstra's shortest path algorithm (\ref{algo3:dijks1}, \ref{algo4:re_dijks2}) twice to extract a clean graph structure of the vasculature (Fig.~\ref{fig-directed}(e)). Finally, we assign directions to the edges of this contracted graph, yielding a directed graph of the whole vasculature, including disconnected vessel regions (see Fig.~\ref{fig-directed}(f)).

% and so on as shown in figure \ref{fig-flowchart}.

% A/V segmentation mask which is used to calculate a crucial bio-marker called Artery/Vein ratio that reveals arteriolar narrowing liked to 

% (Alg.~\ref{algo7:flow_assignment})

% Figure~\ref{fig-directed}.a,b shows this union graph, where the cyan colored nodes are \textit{branch-nodes} ($degree > 2$), and the golden nodes indicate \textit{end-nodes} ($degree==1$). in Ideally, branch nodes should only appear at branching points; however, we can see the union-graph being dense.

%  in union skeleton algorithm (\ref{algo1:union-graph})
 
At this stage, most of the edges of this directed graph are correct, but, due to discretization and other ambiguities in the skeletonization algorithm, some of the crossing nodes will be shifted either upward or downward. These shifts yield spurious \textbf{sink} and \textbf{source} nodes in the directed graph, which are shown as white circles in Fig.~\ref{fig-patches}. To correct these errors, we developed a set of simple \textit{high level graph operations} (HLOs) that we apply to the directed graph. We determine the optimal HLOs by iteratively minimizing a vessel topology cost function (see Eq.~\ref{eq:graph-cost-fn}) over the space of possible graph edits (see Figs.~\ref{hlo-sh-up-dwon} and \ref{hlo-src-dwn}) to yield the final graph.

% Section \ref{susubsec:HLO} list a handful of

% We used two priors to do A/V label propagation;A/V probability obtained from U-Net, directed graph topology obtained by flow-assignment algorithm(\ref{algo7:flow_assignment}). 

This directed topology graph has the potential to serve as the basis for numerous downstream analysis and diagnosis tasks. For instance, in this paper we used this topology to propagate the artery-vein labels we had initially estimated with our second U-net network. By jointly considering the vessel topology and the AV labels---in particular the constraint that downstream vessels should have the same label as upstream vessels---we were able to improve our artery-vein classification, as well as the graph's overall topology. Artery-vein labels are a necessary step for computing the aforementioned artery-vein ratio, which is diagnostically relevant for hypertensive retinopathy, stroke and coronary
artery disease, among other diseases \cite{av_ratio_10.1016/j.amjhyper.2004.10.011, retinal_signs_and_stroke_baker_retinal_2008}. Other diagnostically relevant features that one could extract using our directed graph include branching factors, tortuosity, and crossing anomalies.

% Once we have a low cost topology graph, we simply propagate Artery/Vein labels by using simple propagation strategy discussed in section \ref{subsec:label-prop}. 

The rest of this paper is organized as follows. In Sec.~\ref{sec:literature-review}, we review prior work on retinal vessel segmentation, artery-vein classification, and topology extraction. Then, in Sec.~\ref{sec:methodology} we detail our topology pipeline, and in Sec.~\ref{sec:experiments-and-results} we present experimental results on artery-vein classification using our graphs. For these experiments, we ran multiple ablation studies to determine how the quality of priors (i.e., the initial vessel segmentation and artery-vein labels) affect the quality of the extracted graph, as well as any downstream tasks, such as the artery-vein label propagation. We then discuss these findings in Sec.~\ref{sec:discussion} and discuss future research directions in Sec.~\ref{sec:conclusion-and-future-work}.

\section{Related Work}
\label{sec:literature-review}
Below we review some relevant prior work for retinal graph extraction and artery-vein classification. Most existing work in retinal vessel analysis focuses on generating pixel-wise masks, i.e., where each pixel is labeled either a background or vessel pixel. In the case of AV classification \cite{HEMELINGS2019101636,av_class_vessel_const_10.3389/fcell.2021.659941,multi_task_https://doi.org/10.48550/arxiv.2007.09337,review_10.1007/978-981-10-6496-8_66,XU20173,8999616,Hu2021}, the vessel pixels are sub-classified into either arteries or veins. Some of these approaches achieve good numerical results in terms of accuracy and similar metrics; however, pixel-level masks do not fully solve the medical problem we're trying to address, which is to generate an actionable representation of a patient's vasculature. In particular, the point of estimating the vasculature is to use it to calculate clinically relevant metrics, such as the artery-vein ratio, vessel tortuosity, branching factors, etc., which can then be used by a clinician to determine a diagnosis.

% A handful of graph-based AV classification methods do exist, but they are either limited to the main vessels or require significant manual input. Specifically, the techniques proposed by \cite{7120990} and \cite{tree_morph1019455} requires manual human intervention to correct errors in the graph extraction. Other graph-based techniques rely on color-based clustering and vessel tracking \cite{color_cluster_Vazquez2013} or use a combination of a CNN to classify arteries and veins and a minimum spanning tree to extract the vessel paths \cite{cnn_mst_8309054}. 

% Which, practically is close to impossible from pixel mask because of the imperfections of the pixel level segmentation. For an instance, presence of a small noise in a vessel path could misinterpret the downstream vessel tree totally differently.

% Correct topology extraction, and artery-vein labeling heavily relies on the ability of a model to correctly segment vessels.

Due to the limitations pixel-level masks, a growing number of methods seek to estimate a graph-based topology of the vessels. Fainter vessels and vessels with pathology, in particular, can benefit from topology extraction since the method can leverage global information about blood flow to make up for the lack of local information about these vessel segments. Thus, recent studies have started incorporating the graph as a prior for artery vein segmentation along with image features like color and texture \cite{graphav1_6517259,7120990, graphav_210.1117/12.878712}. These technique rely on detecting branch points and classifying entire segments rather than individual pixels to minimize local errors. Other techniques extract a directed graph, treat the entire vasculature as a sub-graph, and perform label propagation \cite{directed1_7219463,directed2_8008851} or use separate networks for blood vessel, optic disc, and artery-vein labels \cite{top_aware_shin_lee_yun_lee_2020}.

% making it computationally too expensive.

The prior techniques most similar to our proposed pipeline include \cite{hu_automated_2013, hu_automated_2015}, \cite{8754802}, and \cite{Amil2019}, all of which use simple morphological operations, such as skeletonization, to extract a preliminary graph structure. They also identify landmarks, such as branch points, to generate the final graph. Notably, These prior works require a separate method to identify the optic disc; our method, in contrast, identifies this anatomical structure as part of the graph extraction itself. 

More generally, all existing graph extraction methods are (1) either limited to the main vessels or (2) require significant manual input. To the best of our knowledge, our proposed method is the first fully automated graph-based vessel extraction method that can identify the entire vasculature. Below, we discuss our proposed technique in more detail.

\section{Methodology}
\label{sec:methodology}
In this section, we describe the four major steps of our fully automated vessel extraction method: \textbf{\textit{A.}} Vessel and artery-vein prior generation, \textbf{\textit{B.}} Graph extraction with multilevel skeletonization, \textbf{\textit{C.}} Flow estimation with graph-theoretic operations, and \textbf{\textit{D.}} Artery-vein label propagation using the estimated topology. Table~\ref{tab:algo-outline} provides a summary of the different algorithms used in our pipeline, while Tbl.~\ref{tab:symbols} lists the symbols used throughout the rest of this paper. Super- and subscripts are only used when necessary to differentiate between different stages of the algorithm. We describe each of the aforementioned steps below.

\subsection{Vessel and artery-vein priors}
\label{subsection:vessel_av_priors}
% The first step of our pipeline consists of extracting vascular topology and measure different important factors like (Artery to Vein )AV-ratio, turtosity, bifurcation mentioned in the section \ref{sec:introduction}. 

The first step of our pipeline consists of segmenting the vessels at the pixel level and obtaining initial artery-vein labels for each segmented pixel. We use custom U-net deep networks for both tasks, detailed below. Importantly, we do not resize the image before segmenting or labeling the pixels to ensure that we do not introduce artifacts or blur any anatomical features. Fortunately, as our experiments show, our U-net networks can process images of different sizes and quality levels with minimal parameter tuning. Below, we first discuss our vessel segmentation network, then detail our artery-vein classification network.

% can handle images captured using diffe 

% these networks process the original miage Since one of the objectives of our topology estimation is help extract relevant features in the image, we 

% Its crucial to correctly segment vasculature in order to compute usable such measures in the original image, thus our method works on the original image size without having to resize. Funduscope, an imaging device to capture fundus images comes in different size and quality, can have different resolution images and we have shown that our technique works in variety of datasets end-to-end with no to minimum parameter optimization.

\begin{figure*}[ht!]
    \centering
    \includegraphics[width=0.9\textwidth]{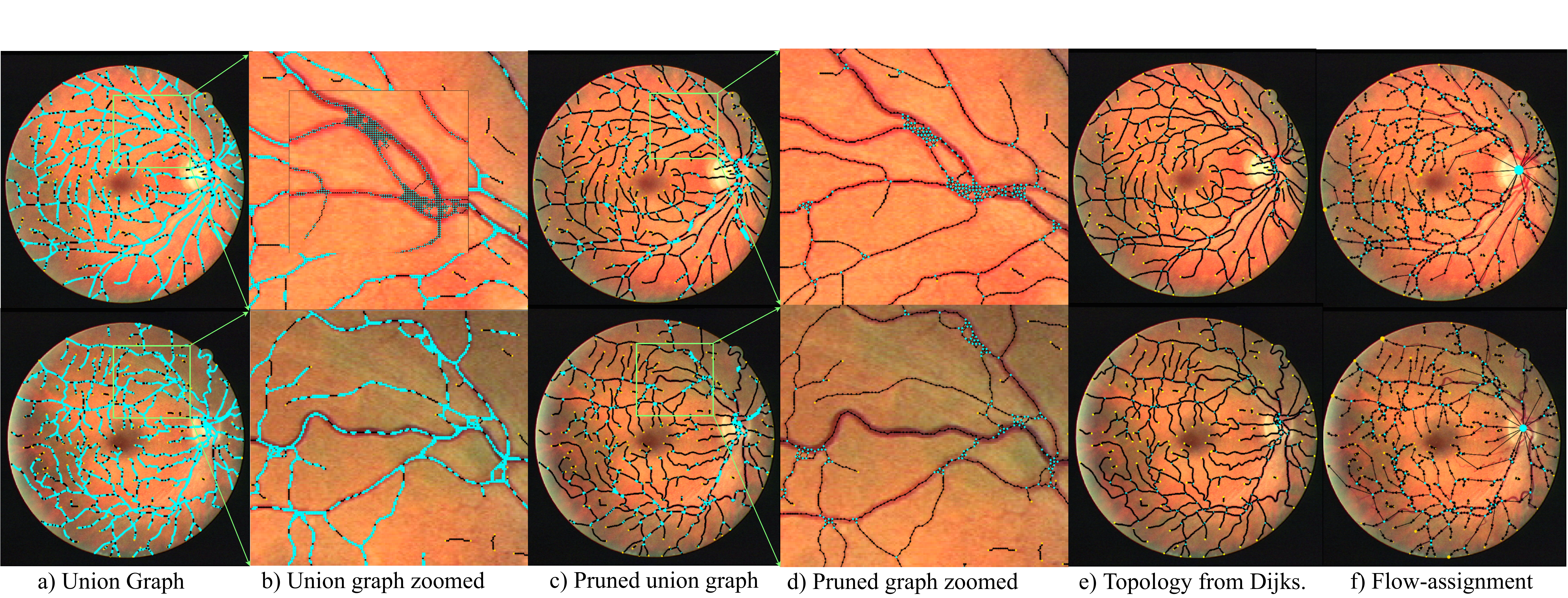}
    \caption{Graph extraction pipeline: (a) Undirected dense graph representing vessel path (Alg.~\ref{algo1:union-graph}). (b) Zoomed union graph to show nodes and edges. (c) Pruned graph with significantly less nodes and edges but with intact vasculature (Alg.~\ref{algo2:union-graph-contraction}) (d) Zoomed pruned graph. (e) Result of applying Dijkstra's shortest path algorithm on the pruned graph (Alg.~\ref{algo3:dijks1} and \ref{algo4:re_dijks2}).  (f) Edge directions assigned using our flow assignment algorithm (Alg.~\ref{algo7:flow_assignment}). For better visualization, we have only shown the median node of each segments and branch nodes. Note that we only use the vessel likelihood map to generate the directed graph. See text for details.}
    \label{fig-directed}
\end{figure*}

% , however for High Level graph Operations(HLO) and label propagation we map the Artery/Vein prior to this final directed graph and go from there.

\subsubsection{Network for vessel likelihood estimation}
\label{sec:likelihood-map-extraction}
% Let $\gls{vessel-pmap}$(fig. \ref{fig-ImagesPrior} c) be the likelihood map of a fundus image $\gls{image}$(fig. \ref{fig-ImagesPrior} a) generated by a trained neural network. 

We use a convolutional neural network on the green channel of the fundus image to estimate a vessel likelihood map, denoted as $\gls{vessel-pmap}$. This map specifies the likelihood that a given pixel is part of a vessel. Specifically, we use a U-Net architecture \cite{Ronneberger2015UNetCN} with a stochastically weighted loss, as introduced in \cite{deeddyn_10.3389/fcomp.2020.00035}, and used successfully by \cite{sam_10.1007/978-3-030-64559-5_41, sam_9680224, dcpa_2110.00512}. A stochastic loss function helps the network detect finer and more ambiguous vessels, leading to a more complete segmentation of the vasculature. The resulting vessel map assigns a continuous probability between 0 (background) and 1 (vessel) to every pixel in the image. As Fig.~\ref{fig-ImagesPrior} shows, though, almost all pixels are assigned values that are very close to one of these two extremes.

% with ease\cite{deeddyn_10.3389/fcomp.2020.00035}.

%  to train k-fold for DRIVE \cite{drive_dataset} and WIDE \cite{wide_dataset:6987362}.
 
\subsubsection{Network for AV-prior estimation}
\label{sec:av-prior-extraction}
We use a U-net network with the same architecture as above, save for the number of outputs (3 vs. 2), to estimate the likelihood that a given pixel is either an artery or a vein. The output of this network, denoted $\gls{av-pmap}$, has three possible labels (artery, vein, or background). As illustrated in Fig.~\ref{fig-ImagesPrior}, this U-net network is capable of correctly classifying most of the vessels in the image, but it suffers from some noticeable errors, such as labeling a branch downstream of a vein an artery. As we detail in Sec.~\ref{subsection:topology-estimation}, we use our extracted graph to refine this initial estimate.

\begin{table}[t]
    \caption{\textbf{Topology Estimation Algorithms:} The above lists the main algorithms used in our graph estimation pipeline. See Fig.~\ref{fig-directed} for a visual example of each of the above steps.\\}
    \begin{tabular}{>{\raggedright}p{2cm}p{5cm}}
        \textbf{Algorithm} & \textbf{Details} \\
        \toprule
        
        \textbf{\ref{algo1:union-graph}}) \textit{Multilevel skeletonization} & Takes a probability map of a fundus image and generates a union graph (Fig.~\ref{fig-directed}(a)). \\
        \midrule
        
        \textbf{\ref{algo2:union-graph-contraction}}) \textit{Union graph contraction} & Uses a graph contraction scheme on the union graph to significantly prune it while still maintaining the overall vascular structure intact (Fig. \ref{fig-directed}(c)). \\ 
        \midrule
        
        \textbf{\ref{algo3:dijks1}}) \textit{Parallel Dijkstra's shortest path vessel tracking} & Takes the contracted graph with weights calculated using heuristics in Eq.~\ref{eq:edge-embedding} and outputs smooth vessel path.\\
        \midrule
        
        \textbf{\ref{algo4:re_dijks2}}) \textit{Dijsktra second pass} & A slightly modified, parallel version of Dijkstra's shortest path algorithm with adjusted edge weights to track vessel in unvisited regions of the graph (Fig. \ref{fig-directed}(e)). \\
        \midrule
        
        \textbf{\ref{algo7:flow_assignment}}) \textit{Flow assignment} & Takes the undirected vessel paths generated above and \textbf{i).} Assigns a direction to each edge using a directed routing strategy (Alg.~\ref{algo6:route}).  \textbf{ii).} Assign pseudo-ONH weights to every end nodes(node with degree 1) and the real ONH to identify disconnected regions in the vascular structure (Fig.~\ref{fig-vonh}).\\
   
        \toprule
    \end{tabular}
    \label{tab:algo-outline}
\end{table}

% depicts the algorithmic steps for topology estimation
 
% because a lot of incoming flow gather in that disconnected point like in

% See fig. \ref{fig-directed}.f for an example.

%  Since Dijkstra's uses local heuristics, it can always take some low cost path leaving some unvisited vascular regions. 
 
%  \textbf{(Alg. \textbf{\ref{algo5:re_dijks2_bw_eight_adjst}})} of the graph used in Algo. \textbf{\ref{algo3:dijks1}}
 
\subsection{Undirected Graph Extraction}
\label{subsec-method-top_extraction}
The second stage of our pipeline consists of estimating an undirected graph given the vessel likelihood map. This extraction has three key steps, detailed below. First, we skeletonize the segmentation at multiple scales, then use Dijkstra's shortest-path algorithm to generate smooth vessel paths starting at the optic nerve. Finally, we clean up this graph to remove self-loops and other small errors arising from the skeletonization. We describe each of these sub-steps below.

% infer whether a given vessel segment is pointed towards or away from the optic nerve.

\begin{table}[ht!]
    \centering
    \footnotesize
    \caption{List of symbols used in the paper}
    \label{tab:symbols}    
    \begin{tabularx}{\linewidth}{lX}
    \textbf{Symbol} & \textbf{Description}\\
    \hline
    $T$ & Thresholds\\
    $I$ & Fundus image\\
    $I_p$ & Vessel segmentation ground truth\\
    $\hat{I}_p$ & Vessel segmentation likelihood (probability map)\\
    $I_{av}$ & Artery-vein ground truth\\
    $\hat{I}_{av}$ & Artery-vein likelihood (prior)\\
    $d$ & Min diameter for connected components to be part of a vessel\\
    $w_v$ & Weights by vessel prob. map in Dijkstra's algorithm\\
    $w_c$ & Weights by pixel color in Lab space in Dijkstra's algorithm\\
    $w_w$ & Weights by pixel distance from background in Dijkstra's algorithm\\
    $\Gamma$ & A vessel path repel factor. How closely two Dijkstra's path can be traced closely. It helps avoid two closely passing vessels touch each other \\
    $L$ & Min number of nodes for a segment\\
    $D_I$ & Number of iterations in fine-tune Dijkstra's algorithm\\
    $k$ & $k$-expansion in fine-tune Dijkstra's algorithm. It represents how much connectivity an un-visited sub-graph has to maintain to the best current topology. \\
    $l$ & False segment limit\\
    $\Delta$ & Cost jump for prop\\
    $r$ & Optic nerve head (ONH) radius\\
    $b_{th}$ & Default branch-forward pair theta coeff.\\
    $a_{th}$ & Default branch-forward pair theta coeff. for A/V Pair\\
    $a_{ce}$ & Labeled branch-forward segments pair A/V cross-entropy coeff.\\
    $a_1$ & Branch cost pair scale factor\\
    $a_2$ & Branch cost prob. scale factor\\
    $a_3$ & Branchness cost scale factor\\
    $a_4$ & Labeled branch-forward graph cost scale factor\\
    $D_g$ & Destination. End nodes(nodes with degree 1) of graph $g$\\
    $D_t$ & Destination. End nodes of graph with smallest threshold\\ 
    $B_g$ & Branch nodes(nodes with degree $>$ 2)\\
    $S^b_i$ & Segment(List of nodes between two branches) starting from a branch $b$ towards its neighbor $i$\\
    $S_g$ & Set of segments in a graph $g$\\
    $N^g_b$ & Adjacent Nodes of $b$ in $g$\\
    $IN^g_b$ & In neighbors of $b$ in $g$, where $g$ is directed\\
    $ON^g_b$ & Out neighbors of $b$ in $g$, where $g$ is directed\\
    $N^i_b$ & List of $i$(integer) neighbors in a lattice of pixel $b$\\
    $H$ & Optic nerve center pixel\\
    $C_g$ & Checkpoint nodes ($D$ + [$H$])\\
    $K^g_n(i)$ & List of $i$ nearest neighbors in $g$ from node $n$\\
    $CC(g)$ & List of connected components of $g$\\
    $G()$ & Empty undirected graph\\
    $DG()$ & Empty directed graph\\
    $Di_g(s, t)$ & Dijkstra's shortest path from $s$ to $t$ in a weighted graph $g$\\
    $w_g(i, j)$ & Weight assigned to edge \textit{(i, j)} in graph $g$\\
    $d_n$ & Degree of node n\\
    $E(v)$ & Average expectation of a vessel (a path of nodes)\\
    $R(v_1, v_2)$ & Artery vein propagation score\\
    $t_{av}$ & Artery-vein propagation score threshold.\\
    \hline
    \end{tabularx}
\end{table}

\subsubsection{Multilevel Skeletonization}
\label{sec:multi-skel-unionG}
The first step to extract a graph given a pixel-level likelihood map is to skeletonize this image. A single skeletonization pass does not yield good results, though, because we need to apply different rules for thick vs. thin vessels. Instead, we use a range of thresholds $\gls{thrs}$ on the likelihood map $\gls{vessel-pmap}$. These thresholds range from $255$ to a minimum $t_0$ (25 in our experiments) with a step size of $p$ (20 in our experiments). In other words, we binarize the segmentation map using each of thresholds, then calculate a skeleton for each threshold, a process we call \textit{multilevel-skeletonization}. We then convert the union of these skeleton into a dense lattice graph, which we call the \textit{union graph} (see Fig.~\ref{fig-directed}(a) for an example). Finally, we use a novel pruning technique to reduce the number of nodes and edges in the graph. Specifically, as shown in Fig.~\ref{fig-directed}(c), we remove small triangles to ensure that all nodes in the middle of a vessel have degree two.

\begin{algorithm}
\SetAlgoLined
$\gls{vessel-pmap}$ = Vessel likelihood map, where each pixel range from $0-255$. \\
$S_T$ = 0 initialized array being same size as the image. \\
% \KwResult Filled $G_u$ \\
\For{t in range(255, $t_0$, -p)}{
    $s_t$ = $threshold(\hat{I_p}, t)$; \\
    $sk_t$ = $skeletonize(s_t)$;\\
    $S_T$ = $max(S_T, sk_t$);\\
}
% $G_u$ = lattice(s_T, conn=8)
\caption{\textbf{Multilevel skeletonization of $\gls{vessel-pmap}$}}
\label{algo1:union-graph}
\end{algorithm}

Algorithm~\ref{algo1:union-graph} summarizes our multilevel skeletonization. In short, the $threshold$ function sets every pixel in $I_p$ to 0 if it is less that the threshold $t$; otherwise it sets it to 255. The $skeletonize$ operation can be any morphological operation that yields a skeleton, and $max$ is pixel-wise maximum. The resulting union graph $G_u$ is an 8-connected lattice graph, where each node is a pixel with value 255 in the thresholded image, and two nodes have an edge if they are 8-neighbor pixel. 

We assign three attributes to each node $(i, j)$ of $G_u$: \textbf{(a)} the vessel likelihood of the corresponding pixel $\gls{vessel-pmap}(i, j)$; \textbf{(b)} the color of pixel $(i, j)$ in Lab space from the original fundus image; and \textbf{(c)} the shortest distance to a background pixel in the binary image with the lowest threshold, which we denote as $bw_{dist}(n)$. We then reduce the number of nodes in $G_u$ using the contraction algorithm detailed in Alg.~\ref{algo2:union-graph-contraction}, which ensures that all nodes that belong to the middle of a vessel segment have degree two. This contraction preserves the topology of the vasculature while significantly reducing the graph size (both nodes and edges), as shown in Fig.~\ref{fig-directed}(c).

%%%%%%%%%%%%%%%%%%%%%%%%%%%%%%%%%%%%%%
% ALGORITHM 
%%%%%%%%%%%%%%%%%%%%%%%%%%%%%%%%%%%%%%
\begin{algorithm}
\SetAlgoLined
$G_u$ = Union graph. \\
$G_{uc}$ = $G_u$.copy() \\
$B_{G_u}^s$ = Branch nodes in $G_u$ sorted in increasing order of vessel likelihood. \\
\While{$B_{G_u}^s$  is not empty}{
    $b$ = pop($B_{G_u}^s$); \\
    \textbf{if} $b$ not in $G_{uc}$: \textbf{continue}; \\
    
    \For{n in $N_b^{G_{uc}} \cap N_b^8$ - $D_t$}{
        \textbf{if} $n$ not in $G_{uc}$: \textbf{continue}; \\
        \textbf{if} $b \ne n_1$: $G_{uc}.addEdge(b, n_1)$; $\forall$ $n_1$ $\in$ $N^{G_{uc}}_n$; \\
        \textbf{if} $n$ in $G_{uc}$: $G_{uc}.removeNode(n)$;\\
    }
}
\caption{\textbf{Union graph contraction}}
\label{algo2:union-graph-contraction}
\end{algorithm}

% $G_u$ = lattice(s_T, conn=8)
% \caption{Union graph($G_u$) contraction to get the lightweight contracted union graph $G_{uc}$}

\subsubsection{Dijkstra's shortest-path algorithm for complete vessel path tracing}

% \textcolor{red}{Before extracting the complete vessel graph, we first detect the optic nerve head (ONH) using the same algorithm but a small trick--we randomly pick a constant pairs(250 in our case) of end nodes and run the Dijkstra's algorithm(\ref{algo3:dijks1}) on them. Then we pick the node with highest intersection on all paths. This works because all vessel must converge to the Optic Nerve Head. If we pick enough source nodes in one side of the ONH, and destinations in the other, the shortest path algo. must pass through ONH in order to reach the destination.}

Even after contraction, the union graph retains some spurious small clusters, as illustrated in Fig.~\ref{fig-directed}(d) (see the groups of blue dots in the image). Thus, we use Dijkstra's shortest path algorithm to refine the vessel paths on this graph further. We run this algorithm in parallel, starting from the different end nodes to reduce the overall running time. We weigh the edges of our graph using the geometric mean of the node attributes mentioned in Sec.~\ref{sec:multi-skel-unionG}:
\begin{equation}
\label{eq:edge-embedding}
w_{(n1, n2)} = e^{w_{v} \cdot log(1+p_*) + w_{c} \cdot log(1+c_*) + w_{w} \cdot log(1+w_*)}
\end{equation}, where 
\begin{itemize}
    \item $p_*$ = 255 - $\frac{I_p(n1) + I_p(n2)}{2}$, and $I_p(n)$ is the vessel likelihood of pixel $n$.
    \item $c_*$ = a color similarity metric, specifically the deltaCie2000(n1, n2) distance in Lab color space \cite{sharma2005ciede2000}.
    \item $w_*$ = $bw_{dist}(n1)$ + $bw_{dist}(n2)$.
\end{itemize}

We first detect the optic nerve head (ONH) by randomly picking pairs of end nodes (250 in our case) and running Dijkstra's algorithm (\ref{algo3:dijks1}) between them. We select the node included most often across all paths as the ONH. We then run two passes of Dijkstra's algorithm starting at the ONH node (see Alg.~\ref{algo3:dijks1} and Alg.~\ref{algo4:re_dijks2}). We run a second pass because Dijkstra's algorithm, by design, consistently favors some paths over others, which leads to some intermediate vessel segments not being utilized at all, which in turn leads to disconnected paths. In order to fix this issue, we first detect unvisited regions, then normalize the edge weights based on the distance to the extracted topology after one pass (Alg.~\ref{algo5:re_dijks2_bw_eight_adjst}). We then rerun the algorithm on these unvisited regions only (Alg.~\ref{algo4:re_dijks2}).

%  rely on greedy approach to be cost effectiveness and could
 
\begin{algorithm}
\SetAlgoLined
$g_0$ = $G()$; \\
\For{c in sorted $CC(G_{uc})$ desc}{
    $c_g$ = $G()$; \\
    src = $H$; \\
    \textbf{if} $H$ not in $c$: src = $D_c$[0]; \\  
    
    \For{parallel t in $D_t \cap c_g.nodes$}{
        $c_{g}.addEdges(Di_{c}(src, t))$; \\
    }
    $w_{c}(i, j)=1.0$; $\forall (i, j) \in$ $c.edges$; \\
    $Q_c= \text{[]}$; \\
    \textbf{if} $len(K^c_n(3))==6$: $Q_c.append(n)$ $\forall$ $n$ $\in$ $c$; \\
    
    \For{parallel t in $Q_c$}{
        $c_{g}.addPath(Di_{c}(src, t))$; \\
    }
    $g_0 = g_0 \cup c_g$; \
}
\caption{\textbf{Parallel Dijkstra's shortest path algorithm for vessel tracking:} This algorithm, which is followed by some fine tuning in Alg.~\ref{algo4:re_dijks2}, takes in the contracted graph from Alg.~\ref{algo2:union-graph-contraction} and outputs a refined, undirected graph.}
\label{algo3:dijks1}
\end{algorithm}

\begin{algorithm}
\SetAlgoLined
$g_1$ = $g_0.copy()$; \\
$g_w$ = $G_{uc}.copy()$ weighted contracted union graph; \\
\For{i in $\gls{dijks2_iters}$}{
   $g_l$ = $g_w.copy()$; \\
   $g_l.removeNodes(g.nodes)$; \\
%   $b_g$ = $branchNodes(g)$; \\
%   $S_g$ = $segments(g)$; \\
   \For{$nodes$ in $CC(g_l)$}{
     $bz = \text{[]} + (N^{gw}_n - nodes)$; $\forall n \in nodes$; \\
     $l_{wg}$ = $g_w.subgraph(K_k^{g_w}(n); \forall n \in nodes \cup bz)$; \\
     $P_{len}[(s, t)] = Di_{g_0}(s, t)$; $\forall (s, t) \in combination(bz, 2)$; \\
     \textbf{if} $len(P_{len}) == 0$: \textbf{continue}; \\
     $P_{len} = sorted_{Desc}(P_{len}.values())$; \\
     
     $bz_{segs}$ = set(); \\
     \For{p in $P_{len}.values()[0]$}{
        \textbf{if} $p$ in $seg$: $bz_{segs}.add(seg)$; $\forall$ $seg$ $\in$ $S_{g_1}$\\
     }
     
     \textit{$BW_{Adjust}$($l_{wg}$, $g_1$}); \\
     \For{seg in $bz_{segs}$}{
        $l_{wg}.edges[e].weight= 0.0$; $\forall e \in$ $l_{wg}.subgraph(seg).edges$\\
     }
     
     \For{b in $b_g \cap l_g.nodes$}{
        $l_{wg}.edges[(b, n)].weight= 0.0$; $\forall n \in$ $N^{l_{wg}}_b$ \\
     }
     
     \textbf{if} $Di_{l_{wg}}^{cost}(P_{len}[0][0], P_{len}[0][1]) > 0.0$: $g.addPath_1(Di_{l_{wg}}(P_{len}[0][0], P_{len}[0][1])$; \\
   }
}
\caption{\textbf{Fine-tune the output of Dijkstra's Alg. \ref{algo3:dijks1}:} The undirected graph extracted by the shortest path topology extraction algorithm is not vessel topology aware. That is, the algorithm always follows the shortest path on a given cost; however, some images can vary in brightness, contrast, or have medical abnormalities leading to disconnected vessel segments. In order to fix this, this algorithm readjusts weights ($BW_{Adjust}$) in such regions where a significant portion of the contracted graph is not visited by Alg.~\ref{algo3:dijks1}. It is parametrized by $\gls{dist_r}$ as in Alg.~\ref{algo5:re_dijks2_bw_eight_adjst}}
\label{algo4:re_dijks2}
\end{algorithm}

\begin{algorithm}
\SetAlgoLined
\SetKwBlock{Fna}{\textnormal{\textbf{$BW_{Adjust}$}$(w_g, g)$\{ }}{}
  \Fna{
    $l_g$ = $w_g$ - $g$; \\
    \For{n in $l_g.nodes$}{
        $w_g.n.dist$ = $dist_{min}(n, b_i)$; \\
    }
    
    \For{$n_1$, $n_2$ in $l_g.edges$}{
        $w_1$ = $255 - \frac{\hat{I}_p(n_1) + \hat{I}_p(n_2)}{2};$ \\
        $w_2$ = $w_g.n_1.dist + w_g.n_2.dist;$ \\
        $w_g(n1, n2).weight$ = $\frac{w_1}{e^{\gls{dist_r} \cdot w_2}};$ \\
    }
  }\}\;
\caption{$BW_{Adjust}$- This algorithm re-initializes edge weights based on how far the left nodes are from the currently extracted topology. The parameter $\gls{dist_r}$ controls how the edge weights are influenced by the distance of two edge nodes from the current topology.}
\label{algo5:re_dijks2_bw_eight_adjst}
\end{algorithm}

\subsubsection{Graph cleaning}
The final step of our undirected graph estimation consists of cleaning any remaining noisy segments, nodes, or edges in the graph. Specifically, our graph cleaning process is as follows:
\begin{itemize}
    \item We use a radius of $\gls{onh_r}$ to contract the optic disk, as shown in Fig.~\ref{fig-directed}(f). We then ignore any nodes withing this disk.
    \item We remove self-loop edges and isolated nodes.
    \item We remove leaf segments with less that $\gls{false_seg_len}$ nodes.
    \item We remove small cycles with $<=$ $\gls{false_seg_len}$ nodes.
    \item We disconnect corner connections (abrupt turns).
    \item We replace very long edges with evenly spaced paths for a more uniform node distribution.
    \item We smooth paths so that they align better with the center of the underlying vessels.
\end{itemize}
We empirically determined the values of the constants listed above for our experiments, which are listed in Tbl.~\ref{tab:exp-parameters}.

\subsection{Directed Topology Estimation}
\label{subsection:topology-estimation}
In this third step of our pipeline, we convert the undirected graph $g$ obtained in the previous step into a directed     graph $d_g$ in which the edges point away from the optic nerve. Intuitively, this corresponds to blood flow for the arteries and the reverse of blood flow for the veins. Estimating a direction for each edge greatly simplifies the topology refinement in subsequent steps since, as proven in \cite{wide_dataset:6987362}, estimating the optimal topology in an undirected graph is NP-hard. Below, we describe our flow assignment algorithm, which is also summarized in Alg.~\ref{algo7:flow_assignment}.

% is robust to assign approximate blood flow.  

\begin{figure*}[ht!]
    \centering
    \includegraphics[width=0.9\textwidth]{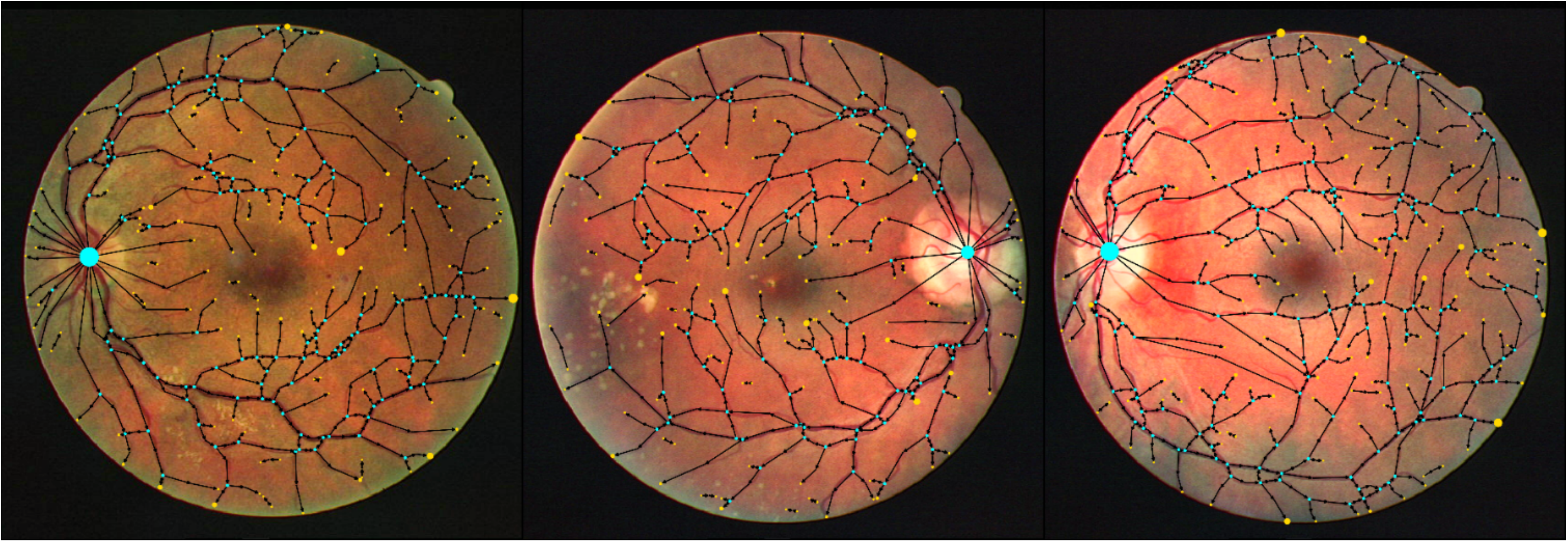}
    \caption{Pseudo ONH obtained by our flow-assignment algorithm (Alg.~\ref{algo7:flow_assignment}) The cyan node denotes the pseudo optic nerve head (ONH), whereas yellow nodes are end nodes (degree = 1). The size of the node indicates the proportion of flows that converge to that node in our flow-assignment algorithm. In short, we traverse from the end points and branch node neighbors until we hit either the ONH or another end node. We can see that most of the visits are accumulated at the pseudo ONH, as well as at the closest end nodes for disconnected vessel segments.}.
    \label{fig-vonh}
\end{figure*}

\subsubsection{Flow assignment}
We first assign a direction to each edge using a recursive flow assignment algorithm. This algorithm takes the undirected graph estimated in Sec.~\ref{subsection:topology-estimation} and determines which end of the edge is pointing away from the optic disc, as illustrated in Fig.~\ref{fig-directed}(f). Correctly estimating the direction of the edges is a crucial step as direction is a key prior for properly estimating the final topology. 

\begin{figure}[ht!]
    \centering
    \includegraphics[width=0.5\textwidth]{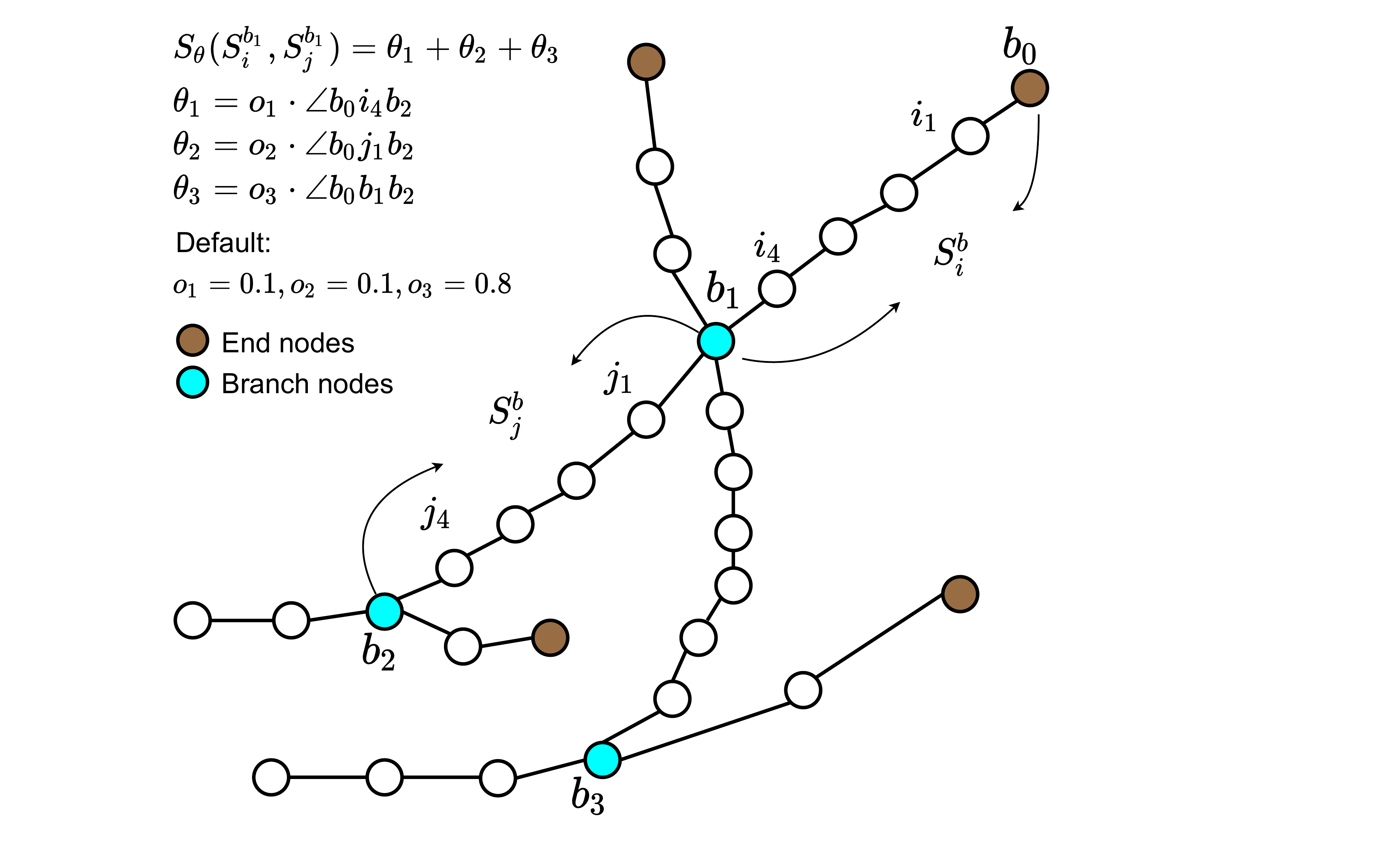}
    \caption{Default heuristics used in the branch-forward measure in Eq.~\ref{eq:fw_default}. This measure is based on the orientation of neighboring branch segments ($S_\theta$). The checkpoint of interest here is $b_0$, and we aim to forward this flow from segment $S_i^b$ through branch $b_1$. The cost of moving along a branch is proportional to how straight the path is.}
    \label{fig:fw_default}
\end{figure}

In more detail, our recursive algorithm determines edge directions based on the geometry of how vessels branch. When a child vessel branches from a parent vessel, it tends to split at an acute angle of $< 90^o$ in a direction away from the optic nerve head (ONH). Our algorithm also uses a momentum factor that models how multiple small flows converge to a larger one, similar to rivers. The core of this algorithm is a \textit{route function} (Alg.~\ref{algo6:route}) that traces vessels from the end nodes ($d_n = 1$) and branch nodes ($B_g$) back to a \textit{checkpoint node} ($C_g$), tallying the number of visits to each checkpoint. These checkpoints are either the ONH or the closest endpoint to the ONH along a path with a missing vessel segment (see Fig.~\ref{fig-directed} for visual examples of these checkpoints). The route function requires a branch-forward measure to select which path to take along a branch. In principle, one can choose a variety of branch-forward measures, but we empirically settled on the following measure:
\begin{equation}
    \label{eq:fw_default}
    fw(g, i, b, j) =  exp(\gls{a_th} \times \frac{180}{1+S_{\theta}(S^b_i, S^b_j)}).
\end{equation} 
Here, $S_\theta$ is the straightness of segment $S_i^b$ when forwarding to another segment $S_j^b$ of a branch $b$ s.t. $j \ne i$. As illustrated in Fig.~\ref{fig:fw_default}, this straightness measure is based on the weighted angles between neighboring segments, parametrized by weighting constants $o_1$, $o_2$, and $o_3$. In short, to determine the direction of flow along a segment, we select the direction that yields the straightest path back to a checkpoint.

% such that small misalignment of neighboring nodes do not lead to incorrect choices. We use the above equation to assign edge directions in our flow assignment algorithm. 

\begin{algorithm}
\SetAlgoLined
$g$ = any un-directed graph. \\
$b$ = any branch node in $g$. \\
$i$ = any adjacent node of $b$. \\
\SetKwBlock{Fna}{\textnormal{path = \textbf{route}$(g, b, i, C_g)$\{ }}{}
  \Fna{
    \textbf{if} $S^b_i\text{[-1]}$ in $C_g$: return $S^b_i$; \\
    $nxt$ = $\min_j(S_\theta(g, i, b, j)) \forall j \in N^g_b - i$; \\
    return $S^b_i\text{[-1]}+ route(g, S^b_i\text{[-1]}, nxt, C_g)$; \\
  }\}\;

\caption{Recursive route estimation used as a subfunction in Alg.~\ref{algo7:flow_assignment}. This algorithm uses a vessel straightness measure (Fig.~\ref{fig:fw_default}) based on vessel orientation. Note that we can replace the branch forward measure with any desired measure.}
\label{algo6:route}
\end{algorithm}

% used in Alg.~\ref{algo7:flow_assignment}, where the branch forward function is given by Eq.~\ref{eq:fw_default}

In healthy vessels, most paths will converge back to the optic nerve head (ONH), as illustrated in Fig.~\ref{fig-vonh}. In this figure, node size corresponds to number of visits to each checkpoints, showing that most of the convergence is to the ONH. Note however, that some paths converge to other checkpoints. These usually reflect disconnected parts of the vasculature where the vessel segmentation algorithm did not detect any vessels, which may be due either to poor imaging conditions or pathologies. As we discuss in Sec.~\ref{sec:discussion}, this distribution of checkpoint convergences could potentially be used as a feature for identifying vascular abnormalities.

\begin{algorithm}
\SetAlgoLined
$g_d$ = $DG()$. \\
$C_v$ = \{$c_i:0$\} $\forall$ $c_i$ $\in$ $C_{g_1}$. \\
\For{$n$ in $D_{g_1}$ + $B_{g_1}$}{
    \For{$nbr$ in $N^{g_1}_n$}{
        $pth$ = $route(g_1, n, nbr, C_{g_1})$; \\
        $C_v[pth[-1]] \text{ += } 1$;
    }
    \For{$seg$ in $S_{g_1}$}{
        $f_1$ = $route(g_1, seg[0],seg[1], chk)$; \\
        $f_2$ = $route(g_1, seg[-1],seg[-2], chk)$; \\
        $S_{f1}$ = $C_v[f_1[-1]] \cdot (1-int(seg[-1] \in ends))$ + $\frac{1}{len(f_1)}$; \\
        $S_{f2}$ = $C_v[f_2[-1]] \cdot (1-int(seg[0] \in ends))$ + $\frac{1}{len(f_2)}$; \\
        
        \textbf{if} $f_1 > f_2$: $seg$ = $seg[::-1]$; \\
        $g_d.addEdges((seg[i], seg[i+1])$; \\
    }
}
\caption{Flow assignment: This algorithm takes as input an undirected planar graph $g_1$. It traverses a path back to a checkpoint ($C_{g_1}$), either the ONH or the closest end-point to the ONH, starting at each end node ($D_{g_1}$, $d_n=1$) and branch node ($B_{g_1}$) using the routing algorithm described in Alg.~\ref{algo6:route}. In addition to the edge directions, we also tally the number of hits to each checkpoint.}
\label{algo7:flow_assignment}
\end{algorithm}

\begin{figure*}[ht!]
    \centering
    \includegraphics[width=0.9\textwidth]{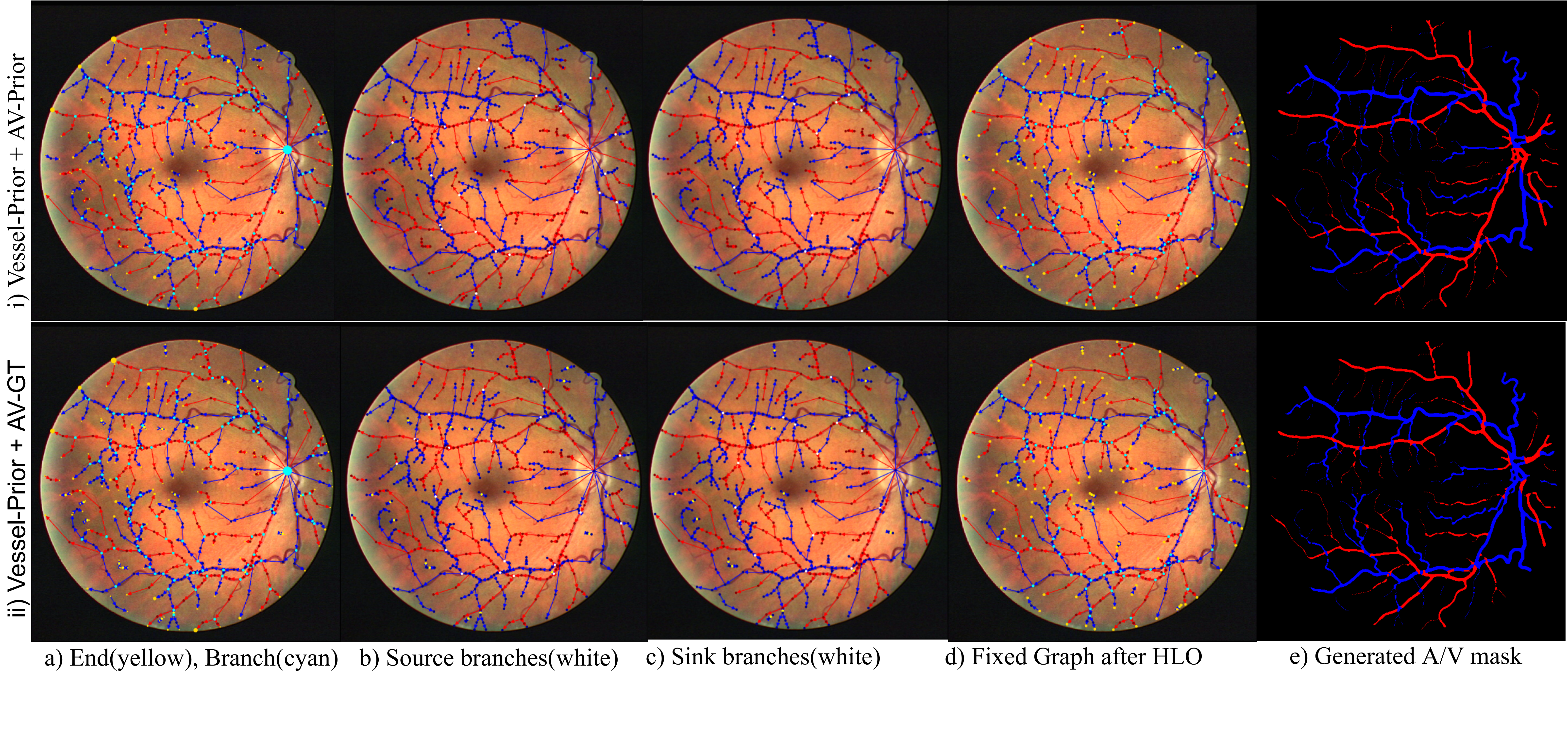}
    \caption{Prior Mapped to flow-graph obtained from Alg.~\ref{algo7:flow_assignment}. i) Flow obtained from vessel prior(from U-Net), and AV-Prior(U-Net). ii) Flow obtained from vessel-Prior and AV-Ground Truth. a) End nodes(golden yellow, $degree=1$), branch nodes(cyan, $degree>2$. b) Source branch nodes(white), c) Sink branch nodes(white), d) Graph after High Level Graph Operations(HLO) + A/V label propagation. e) Artery/Vein segmentation mask generated from propagated topology graph by mapping it to a binary segmentation and using nearest neighbour label propagation.}
    \label{fig-22-test-Mappings}
\end{figure*}

\subsection{Tree Topology Estimation from Directed Graph}
After assigning a direction to each edge, the resulting graph $d_g$ is, by construction, a directed \textit{planar} graph. However, vessels only appear planar in a fundus image because the arteries and veins cross each other, creating spurious crossing points in the image. Therefore, in the final stage of our pipeline, we convert the planar topology into a three-dimensional graph by splitting these crossing points in a way that best matches the underlying anatomy. In other words, we seek the most likely set of splits such that the resulting graph consists of separate artery and vein subtrees (i.e., subgraphs without loops). Below, we first discuss the cost function that we use to determine which tree topologies are most likely given the planar graph, then detail the high-level operations (HLOs) that we use to merge and shift nodes in the graph to obtain a better overall topology.

\subsubsection{Tree Topology Cost Function}
Formally, we seek to convert a directed graph with loops into a tree without loops. This tree should have two main subtrees, one for the arteries and one for the veins. We convert a directed graph into a tree by splitting the crossing nodes, i.e., nodes with in-degree greater than one, into multiple nodes, each with in-degree equal to one. To determine how to split these nodes, we use the following cost function to assign a likelihood to each possible tree topology:
\begin{equation}
    \label{eq:graph-cost-fn}
    C_{g_d} = \sum_{b \in B_g} e^{CR_{pair}(b)} + e^{CR_{prob}(b)} + e^{BR(b)} + e^{FW(b)}.
\end{equation}
Here, $g$ is the undirected graph of $g_d$,  $CR_{pair}$ is a \textit{crossings cost} that measures if the number of incoming arteries and veins match the number of outgoing arteries and veins at crossing nodes. $CR_{prob}$ penalizes segments that have both artery and vein labels along their path, while $BR$ penalizes crossings with unnatural branching (e.g., when two incoming vessels have only one outgoing vessel). Finally, $FW$ penalizes artery-artery and vein-vein crossings. We discuss each term in more detail below.

% cost keeps the blood flow between A-A/V-V in check.
The first term in the cost function, $CR_{pair}$, is defined as:
\begin{equation}
    \label{eq:GC_CR-Pair}
    CR_{pair}(b) = {a_1} \cdot c_b, \\
    c_b = \begin{cases} 3-max(A_{c}, V_{c}), \text{if $d_b = 3$} \\
                                        \frac{|A_{c}-V_{c}|}{cr_{div}}, \text{otherwise} \end{cases}
\end{equation}
where $A_c$ is the number of artery segments connected to branch $b$, $D_b$ is the degree of branch $b$, and $cr_{div}$ is a normalization factor that enforces that the number of arteries and veins in a crossing should be the same when the degree of that crossing is 4, as well as that the arteries and veins should be paired as much as possible for higher degrees. We use $cr_{div}=1.9$ (instead of 2) to ensure that we do not penalize natural $A-V$ crossing for cases where $D_b > 4$.

Our second term, $CR_{prob}$, is defined as:
\begin{equation}
    \label{eq:GC_CR-Prob}
    CR_{prob}(b) = a_2 \cdot (1-|E(a) - E(v)|),
\end{equation}
where $E(a)$ is the average likelihood that nodes in neighboring segments of branch $b$ are arteries, and likewise $E(v)$ for veins. This term enforces that the likelihood of segments along a branch being arteries or veins should be nearly discrete (i.e., close to either 0 or 1).

% Any violation to this triggers both $CR_{pair}$ and $CR_{prob}$.

Similarly, $BR$, detailed in Alg.~\ref{eq:GC-BR}, is the cost incurred when unlikely branching scenarios occur---such as an incoming vessel in a crossing having no outgoing vessel. That is, the \textit{out-degree} should ideally be greater or equal to the \textit{in-degree}. Also, when there are multiple vessels crossing at a single point, the segments that share a minimum branch-forward cost ($S_\theta$), as illustrated in Fig.~\ref{fig:fw_default}, should be preferred.

\begin{algorithm}
\SetAlgoLined
\SetKwBlock{Fna}{\textnormal{\textit{BR}$(g, b)$\{ }}{}
  \Fna{
    $gc =[\gls{eps}]$; \\
    $x = max(S_{\theta}(S_i^b, S_j^b))$; \\
    \textbf{if} $d^{in}_b >= 2$\textbf{:} $gc.append(\frac{x}{360}); \forall i, j \in IN_b^g$; \\
    \textbf{if} $d^{out}_b >= 2$\textbf{:} $gc.append(\frac{x}{360}); \forall i, j \in ON_b^g$; \\
    return $\gls{a_3} \cdot [((\frac{d^{in}_b}{d^{out}_b + \gls{eps}})^2 + \frac{\sum{gc}}{len(gc)})]$;
  }\}\;
\caption{Branchness cost for each branch in a directed graph $g$, where $d^{in}_b/d^{out}_b$ is the in/out degree at branch $b$ of the corresponding directed graph $g_d$.}
\label{eq:GC-BR}
\end{algorithm}

% aligns with the artery-vein prior discussed in Sec.~\ref{subsection:vessel_av_priors}

Finally, the $FW$ term measures if the branch-forward operation in Alg.~\ref{eq:GC-BR} results in labels that are consistent with the way arteries and veins are distributed in a retinal image. That is, this term penalizes neighboring segments having inconsistent labels (e.g., one is labeled an artery, the other a vein). To measure this consistency, we use a version of the \textit{branch-forward} cost that has the artery-vein prior factor added along with the orientation (straightness):
\begin{equation}
    \label{eq:fw_av}
    S^{av}_\theta(i, b, j) = e^{\gls{a_th} \cdot log(1+S_\theta(g, i, b, j)) + \gls{a_ce} \cdot log(1+ce(S^b_i, S^b_j))},
\end{equation}

\begin{equation}
    \label{eq:GC-FW}
    FW(b) = \gls{a_4} \cdot \frac{
    \sum_{i \in IN_b^{g_d}}
    {S^{av}_\theta(i, b, S^{n_{min}}_\theta(b, i))}
    }
    {
    len(IN_b^{g_d}),
    },
\end{equation}
where $S^{n_{min}}_\theta(b, i) = j,$ s.t. $S_\theta(i, b, j), \forall j \in N^g_b - \{i\}$. In short, a high cost is incurred if, for all the \textit{in-neighbors} of a branch, the straightest outgoing segment has a different label to the incoming segment.

%  as mentioned in fig. \ref{fig:fw_default} as follows

\subsubsection{High Level Graph Operations (HGO) for optimal blood flow estimation}
\label{susubsec:HLO}
In this section, we describe the graph edit operations that we use to refine our estimated topology. As noted above, graph topology estimation is an NP-hard problem \cite{wide_dataset:6987362}. However, in practice the optimization problem is tractable given a good directed graph $g_d$, and a high-quality artery-vein prior (see Sec.~\ref{sec:likelihood-map-extraction}); both of these pieces of information help constrain the overall optimization problem.

\begin{figure}[ht!]
    \centering
    \includegraphics[width=0.5\textwidth]{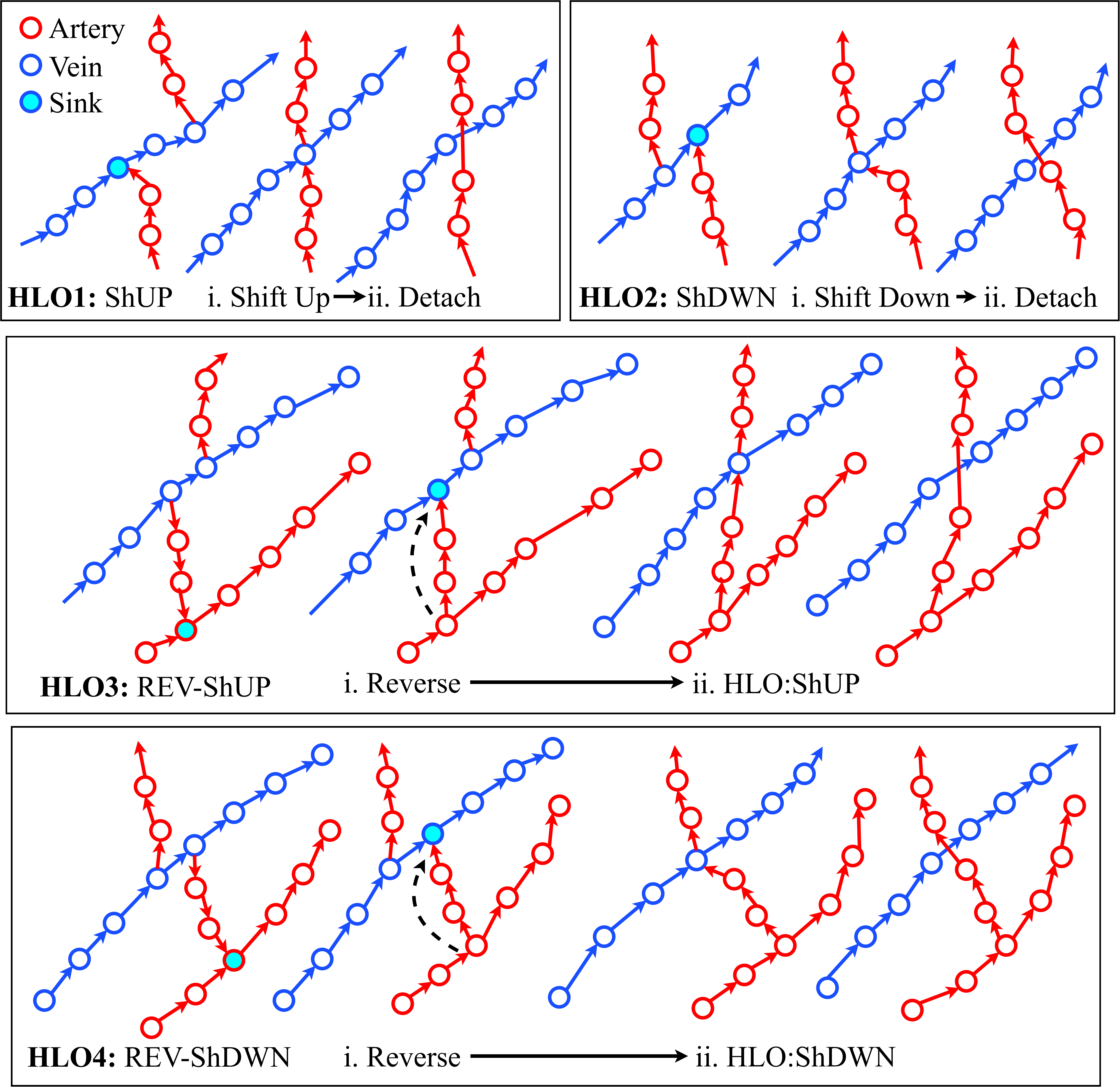}
    \caption{High Level Operation Shift up and Shift Down from sink. We use HLOs to correct errors in the initial skeletonization and graph extraction. See text for details.}
    \label{hlo-sh-up-dwon}
\end{figure}

\begin{figure}[ht!]
    \centering
    \includegraphics[width=0.5\textwidth]{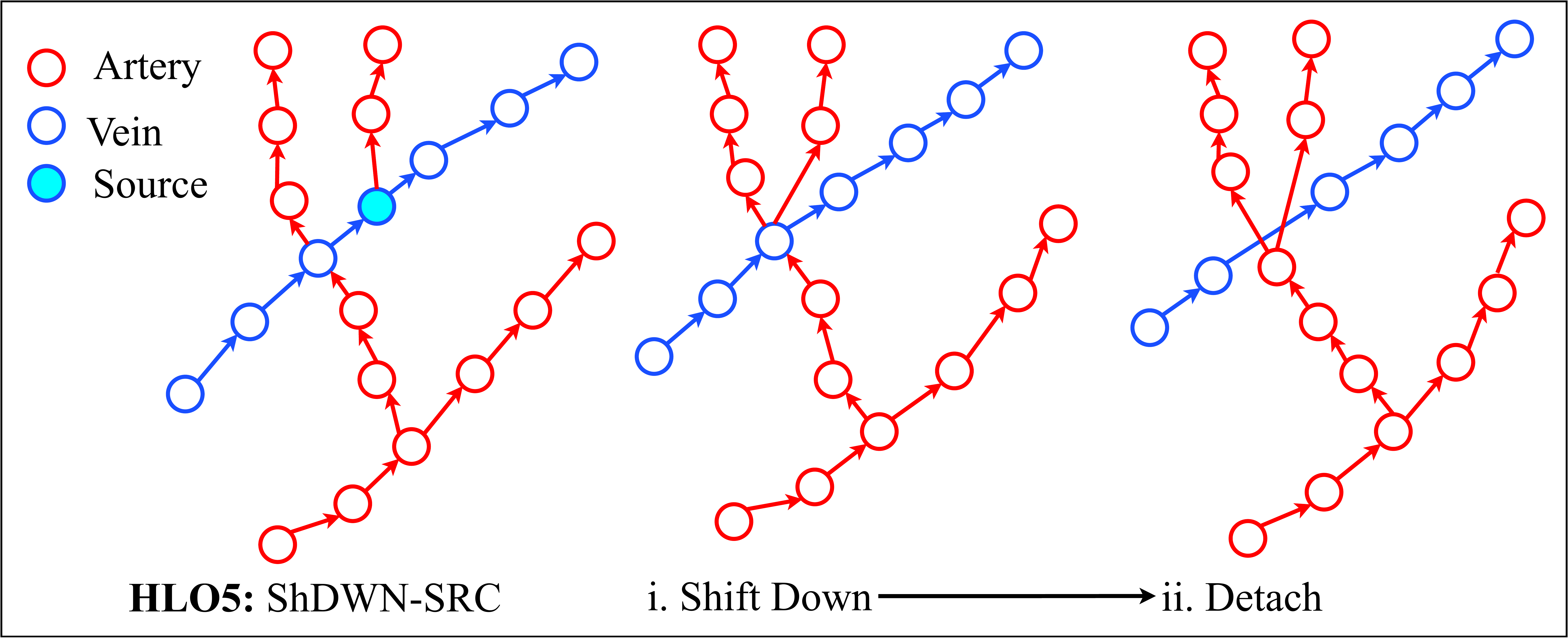}
    \caption{High Level Operation Shift Down from source. We use HLOs to correct errors in the initial skeletonization and graph extraction. See text for details.}
    \label{hlo-src-dwn}
\end{figure}

As such, we first map our artery-vein prior $\gls{av-pmap}$ to the directed graph $g_d$ by assigning to each edge the most common AV label of the pixels that it overlaps (see Fig.~\ref{fig-22-test-Mappings}). We then minimize Eq.~\ref{eq:graph-cost-fn} using a finite set of \textit{high-level graph operations} (HLOs), illustrated in Figs.~\ref{hlo-sh-up-dwon} and \ref{hlo-src-dwn}. We describe each HLO below.

First, however, it is important to understand why graph editing is needed in the first place. For example, when an artery and a vein cross each other, most path extraction algorithm, including ours, tend to shift the crossing either up or down creating two three-degree nodes instead of a single four-degree branch node (see Fig.~\ref{hlo-src-dwn}(HLO1)); thus, one of our HLO operations is tasked with correcting this particular error. Our other HLOs follow a similar pattern: they are meant to correct local errors in the topology stemming from extraction errors or pixelization ambiguity.

In more detail, we use the directed topology graph $g_d$ mapped with A/V labels from $\gls{av-pmap}$ to identify \textit{source} and \textit{sink} nodes (i.e., those with either no in-neighbors or out-neighbors) since we empirically determined that most editing errors involve these types of nodes. Specifically, sink and source nodes are often the result of (1) a branch connecting to the wrong node on the overlapping branch or (2) a segment that is pointing in the wrong direction. We iteratively perform HLOs on these nodes, maintaining a heap of edited graphs with their associated graph cost (Eq.~\ref{eq:graph-cost-fn}), with the lowest-cost topology at the top of the heap. However, we only modify the current topology to fit the top of the heap if and only if the cost difference with the current best state is more than $\gls{c_jump}$. This threshold prevents the optimization algorithm from over-editing the graph. After each HLO, we propagate labels from the shift destination nodes (see Fig.~\ref{hlo-sh-up-dwon}) before calculating the new graph cost since the HLO might cause global changes in the artery-vein labels. We then push the topology onto the heap and select a new possible topology to explore. Below, we discuss how we identify sink and source nodes in more detail.

% This makes sure that our operation is a valid one locally and in the global vasculature context.

% The topology cost, though, is only affected by the local changes from the HLO, not other downstream effects, so thus won't detect any misconnection far off from the source/sink of operation, so each time

% However, if any valid operations are missed because of weak prior, the algorithm being iterative, will eventually propagate to make the necessary jump.

% avoids any noise and only keep robust and correct operations in the optimization path.

\begin{figure}[ht!]
    \centering
    \includegraphics[width=0.5\textwidth]{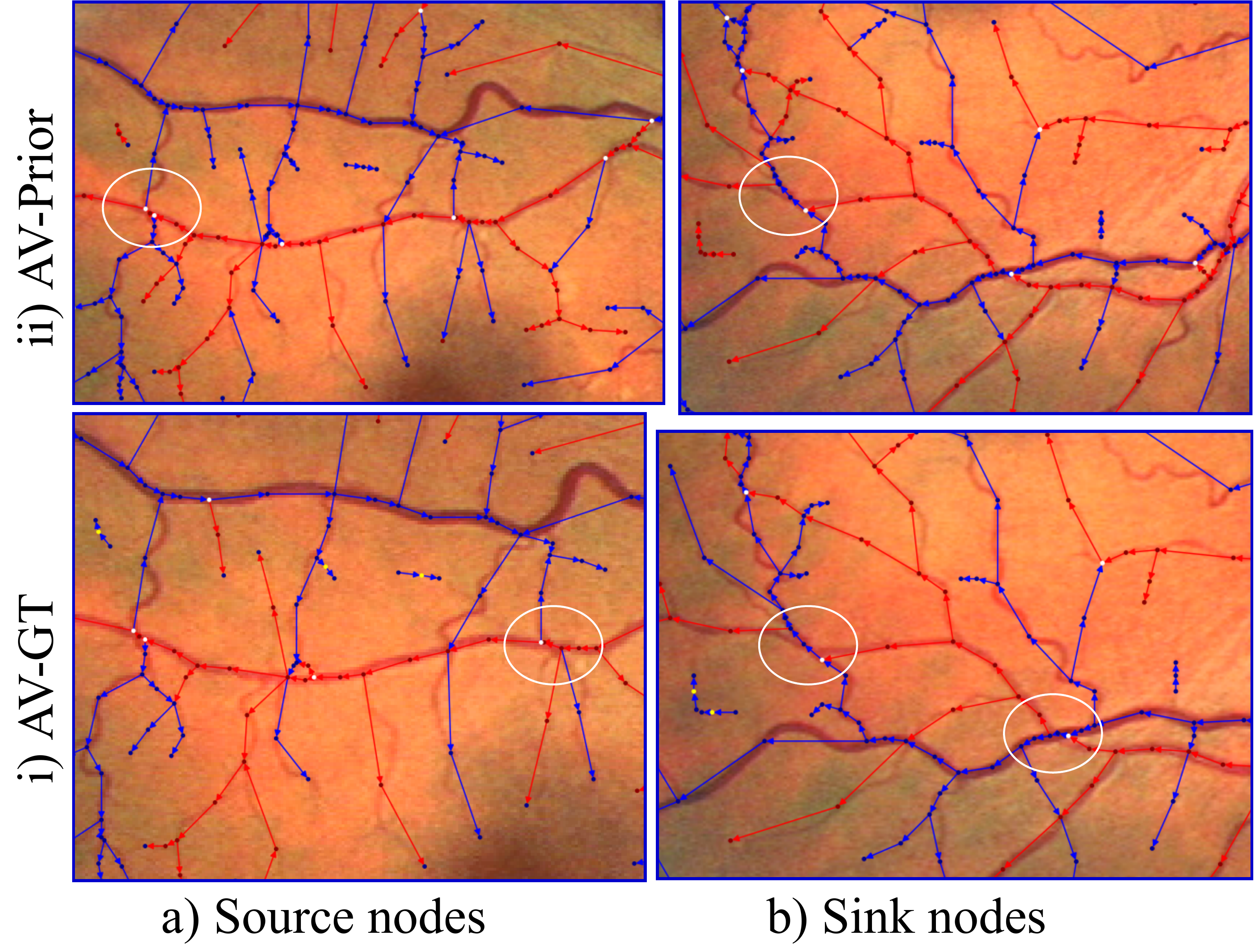}
    \caption{Source and sink nodes in actual fundus image patches. Note how the extraction errors match the HLOs described in the text.}
    \label{fig-patches}
\end{figure}

\vspace{0.5em}
\noindent \textbf{Source nodes:} We define a source node as a branching node where the outgoing vessels have different labels (see (Figs.~\ref{hlo-src-dwn}, \ref{fig-patches}(a)). Arteries and veins can only overlap in the image, not bifurcate from each other, but the AV prior may contain such inconsistencies. To identify these points, we first use the route method (Alg.~\ref{algo6:route}) on the undirected graph $g$ of $g_d$ at two outgoing vessel with the $A/V$-aware forward algorithm $S_\theta^{av}$ (Eq.~\ref{eq:fw_av}). We then calculate a propagation score $R(v_1, v_2)$ (Eq.~\ref{eq:prop-score}) based on the outgoing vessel segments:
\begin{equation}
\label{eq:prop-score}
    R(v_1, v_2, k) = max(\frac{e^{m_a}}{(e^{m_a} + e^{m_v})}, \frac{e^{m_v}}{(e^{m_a} + e^{m_v})})
\end{equation}, where $m_a = z \cdot E(v_1^{artery}) + 1 - E(v_2^{artery})$, $m_b = z \cdot E(v_1^{vein}) + 1 - E(v_2^{vein})$, and $z$ is a scaling factor. We only select a branching node as a source node if the propagation score is greater than a threshold \gls{prop_score_thr}. In addition to this threshold-based selection, we also consider a bifurcated branch node to be a source if the two outgoing sub-graph intersect further downstream. The latter is a necessary constraint because, as mentioned above, arteries and veins never cross each other, thus if the two paths meet, at least one of them has the wrong label. 

% Even if it is because of wrong prior, it is actually better to mark this node as a source because our optimization iteration labels more confident cases first, thus improving the prior to make better decisions. 

% both of them has to be different labels.

%  instead of the default $S_\theta$ described (fig. \ref{fig:fw_default}).

% To avoid propagation that violates this fact

% These are the branch nodes that are bifurcated in the directed graph $g_d$ (Fig.~\ref{hlo-src-dwn}, \ref{fig-patches}(a)). To identify these points, we use the route method (Alg.~\ref{algo6:route}) on un-directed graph $g$ of $g_d$ at two outgoing vessel with the $A/V$ flavoured forward algorithm $S_\theta^{av}$(eq. \ref{eq:fw_av}) instead of default $S_\theta$(fig. \ref{fig:fw_default}).

\vspace{0.5em}
\noindent \textbf{Sink nodes:} Similar to source nodes, a sink node is a node where the two or more incoming vessels have different labels (see Figs. \ref{hlo-sh-up-dwon}, \ref{fig-patches}(b)). We also use the propagation score and threshold discussed above to select these nodes.

% Sink nodes are branch points in directed graph $g_d$ where two vessels converge to a node (Fig. \ref{hlo-sh-up-dwon}, \ref{fig-patches}.b).

% as it makes sure we work on confident propagation first, improving expectation on rest of ambiguous decisions. 

\vspace{0.5em}
\noindent \textbf{High Level Operations:} We now describe our HLOs on sink and source nodes. There are four operations on sink nodes as shown in Fig.~\ref{hlo-sh-up-dwon}: Shift-Up, Shift-Down, Reverse-Shift-Up, and Reverse-Shift-Down. For the Reverse HLOs, one needs to choose which of the two segments to reverse. Here, we pick the segment with the highest $CR_{pair}(b)$ cost (Eq.~\ref{eq:GC_CR-Pair}). Source nodes have two operation. The first is Shift-Source-Up, as shown in Fig.~\ref{hlo-src-dwn}. The second is Shift-Source-Up, which we found to be very rare in our dataset, so we omitted it in our experiments for simplicity. One can easily include this operation as well, though, if needed for a new dataset.

% because there are two choices(two inward vessels in the sink node, fig. \ref{hlo-sh-up-dwon}.3,4 i). A close inspection reveals that
 
\vspace{0.5em}
\noindent \textbf{High Level Graph Operation on Sink Nodes:} We perform the aforementioned four operations on each sink node as follows. First, we sort the sink nodes in ascending order by their distance to the ONH and push each possible topology into a min heap---the top of the heap is the topology with minimum cost. Working with sink nodes that are closest to the ONH first makes more sense because sink nodes that are further away must route through the closer nodes to get to the ONH. Thus, sink nodes closer to the ONH have a higher impact on the correctness of the overall topology. As an example, we can compare the initial sink nodes in Fig.~\ref{fig-22-test-Mappings}(c) vs. after fixing via an HLO operation in Fig.~\ref{fig-22-test-Mappings}(d). In more detail, the four operations, illustrated in Fig.~\ref{hlo-sh-up-dwon}, are:
\begin{itemize}
    \item \textbf{Shift-Up:} We merge the outgoing segment from the sink down to the middle node of that segment. This in turn squeezes the outgoing segment and its end branches to a single branch. We call this a shift-up operation because we shift the sink upwards in the direction of the flow. 
    
    % Also, we start working from the sink that is near to the optic nerve head.

    \item \textbf{Shift-Down:} The shift down operation is similar to shift up but in the reverse direction (opposite the flow).

    \item \textbf{Reverse (followed by shift up or down):} The performance of the flow assignment algorithm (Alg.~\ref{algo7:flow_assignment}) depends on how good the vessel prior $\gls{vessel-pmap}$ is. There could be cases where a segment is assigned the wrong direction because of a wrong prior. To correct this, we use a \textit{reverse} operation, followed by a shift up or down to connect the reversed segment to the correct downstream vessel. 
    
    % It should be noted that not all HLO operation lead to a good optima. The core of such operation is fueled by our graph cost measure in eq. \ref{eq:graph-cost-fn}.
\end{itemize}
% Note that a sink node must be accompanied by a source node in order to make a correct shift UP/DWN operation. 

\vspace{0.5em}
\noindent \textbf{High Level Graph Operation on source:} As mentioned above, source and sink nodes come in pairs in the majority of cases, corresponding to the beginning and end of a vessel segment. However, there are cases where source node appears by themselves; usually such locations form when a vessel passes through the branching of another vessel (see Fig.~\ref{hlo-src-dwn}(HLO5) for an example). In theory, we can apply both shift up and shift down operations in such locations, but our close inspection of the datasets used in our experiments revealed a handful cases of shift down, and almost negligible cases of shift up operations. Therefore, we have only used shift down operations for source nodes. 

% Unlike sink node, source nodes are always worked with from the ones furthest from the ONH for the similar reason as sink nodes.

\subsection{AV Label Propagation}
\label{subsec:label-prop}
When correcting the topology with HLO operations, it is important to update the artery-vein labels so that they match the corrected topology. We update these labels using \textit{label propagation}, namely assigning downstream segments the label of upstream segments. As we show in our experiments (Sec. \ref{sec:experiments-and-results}), this label propagation consistently improves the overall artery-vein classification accuracy across multiple datasets.

% as shown in Experiments and Results section (Sec. \ref{sec:experiments-and-results}). 

% We have explained the vasculature extraction and estimation process in sections \ref{subsec-method-top_extraction} and \ref{subsection:topology-estimation}. In order to validate the claim that topology extracted from our technique is correct, we begin with \gls{av-pmap} and the directed graph ($g_d$, and $g$ for corresponding undirected graph) extracted from section \ref{subsec-method-top_extraction}, and use a simple iterative propagation algorithm to propagate A/V labels correctly. As such, our results have consistently shown improvement in A/V scores as shown in Experiments and Results section (Sec. \ref{sec:experiments-and-results}). 

In more detail, we propagate the labels starting at two points:
\begin{itemize}
    \item \textbf{Sink node propagation:} We start propagating from the sink nodes that are closest to the optic nerve head (ONH). This minimizes the probability of an incorrect propagation because all the paths we are about to propagate labels to have already been worked with. We set the artery-vein probabilities of each node to be the average expectation of the incoming paths obtained by the \textit{route} algorithm (Alg.~\ref{algo6:route}). Once we are done with all sink nodes, we continue iteratively for the remaining nodes until the graph cost in Eq.~\ref{eq:graph-cost-fn} stops improving.

    \item \textbf{Propagate from end nodes}. After propagating from the sink nodes, we propagate labels starting from the end nodes back to the ONH. We use the same routing method (Alg.~\ref{algo6:route}) starting from each end node, replacing each path by the average expectation of the vessel. The average expectation works well in this case because the A/V priors estimated by our neural network are nearly discrete (i.e., always close to 0 or 1). We run multiple passes over all end nodes and stop when the graph cost stops improving.
\end{itemize}

As noted above, this propagation scheme consistently improved the overall accuracy of artery-vein labels. However, it is important to note that one of the challenges of graph-based label propagation is that a single incorrect label can affect all downstream segments stemming from it. For example, a single incorrect label near the ONH could cause an entire subtree to have the wrong label. Thus, we regularize our label propagation using \textit{stop-points}---nodes from which we avoid any further propagation. A good examples of a stop point is a crossing where one of the neighboring segments is undetectable because of poor image quality or a pathology, causing this four-neighbor branch (crossing) to appear as a three-neighbor branch (source node). If one of the neighbor has a different label, it might incorrectly be propagated to the other segments. Another example is when two directed paths from two source nodes meet further down the tree, violating the AV crossing constraint. More generally, we select stop nodes among the source nodes whose outgoing sub-trees meet at some downstream node. This ensures that the two sub-trees are of different labels. We avoid propagating labels beyond such stop-points by adding these stop-nodes to the checkpoint nodes of the corresponding graphs ($C_g$) in the routing algorithm (Alg.~\ref{algo6:route}). 

% we propagate from two places in code. From sink nodes, starting from the ones nearest to ONH. And from end nodes(nodes with degree 1). When propagating from sink nodes, we select stop nodes among the source nodes whose two outgoing sub-trees meet at some point. This makes sure that the two sub-tree indeed are of different labels. Because, in the first place, enough evidence was used to mark it as a source, now an intersection confirms that it is a true case.

\section{Experiments and Results}
\label{sec:experiments-and-results}
We tested our graph estimation and artery-vein classification pipeline on multiple retinal datasets. Specifically, we tested our pipeline on the DRIVE \cite{drive_dataset}, WIDE \cite{wide_dataset:6987362}, and INSPIRE \cite{inspire_dataset-tang2011robust} datasets, which have 40, 30, and 40 images, resp. The first two datasets had ground-truth, pixel-level vessel segmentations and AV labels, while the latter only had sparse AV-labeled graphs specifying the vascular topology. For each dataset, in addition to our full pipeline, we also carried out ablation studies to understand the impact of different stages of the pipeline on the final result. Below, we first describe our experimental setup in more detail, then discuss our quantitative results.  

\subsection{Experimental setup}

\noindent\textbf{Convolutional neural networks:}
As described in Sec.~\ref{sec:av-prior-extraction}, we used two separate U-Net \cite{Ronneberger2015UNetCN} networks to generate pixel-level vessel and artery-vein priors. We trained separate networks for the DRIVE and WIDE datasets using their corresponding ground-truth annotations. For INSPIRE, however, which does not have pixel-level ground truth, we obtained likelihood maps by training a third pair of networks on the combined data from other datasets, described in more detail below. As we show in our results, our pipeline was able to achieve good transfer learning results on this third dataset, confirming the generality of our proposed approach.

% In tables \

% (plus some adjustments like resizing, normalization)

% All neural network related computations

\vspace{0.5em}
\noindent \textbf{Hardware:} We performed all of our experiments using a Dell Precision 7920R server with two Intel Xeon Silver 4110 CPUs (32 threads each), 128 GBs of RAM, and two 1080 Ti Nvidia GeForce GTX graphics cards.

\vspace{0.5em}
\noindent \textbf{Neural network training:} We used 5-fold cross validation to generate likelihood masks for all the images in the datasets. We trained each network for 300 epochs with a patience of 50 epochs until no there was no improvement in the validation set. We used an Adam optimizer with a learning rate of 0.001 and a batch size of 4 for all experiments. We utilized our custom-built PyTorch library, EasyTorch \cite{easytorch}, to train these networks.

% to for facilitating seamless training and evaluation for transfer learning tasks.

\vspace{0.5em}
\noindent \textbf{Graph computations:}
We ran the parallel Dijktra algorithms (Algs.~\ref{algo3:dijks1} and \ref{algo4:re_dijks2}) and all the other graph algorithms using eight CPU threads. We used a separate validation set to optimize the graph-operation parameters (listed in Tbl.~\ref{tab:exp-parameters}), which we then used for all experiments across the three datasets.

\vspace{0.5em}
\noindent \textbf{Pipeline ablation:} In addition to our full pipeline, we tested two ablated versions to understand the impact of both the pixel-level segmentation and the initial artery-vein labels on our final results. In other words, our graph estimation pipeline requires two forms of prior information: (1) vessel segmentation (i.e., which pixels are part of a vessel) and (2) the pixel-level AV labels. In this paper, we used U-net networks for this task, but our pipeline is agnostic as to how these likelihood maps are generated. Ideally, as better deep learning methods are developed for these two tasks, our graph pipeline should automatically improve its results, in turn. To determine the ceiling performance for our graph pipeline, we ran alternate versions of our pipeline in which we replaced the U-net's outputs with the ground truth data, either for the vessel segmentation or the AV labels. As we detail in our results below, our graph pipeline achieved even better results when given ground-truth data, validating that our graph extraction, optimization, and label propagation model the topology of the underlying vessels well.

% and used them for the rest of the data to report the performance.

% It reveals that this technique performs much better if we introduce good priors. In each row group, boldface rows(superscript \textit{*}) is the scores after we perform all proposed techniques, whereas the non-bold ones(superscript \textit{0}) represent the starting state. Table \ref{tab:exp-parameters} shows parameters used in these experiments for each datasets. We explain each of the evaluation mechanism in later sections of this section.

% prior is the  probability maps generated from a trained U-Net model, and subscript \textit{gt} means the respective ground truth; We have substitute one of the priors with the actual ground truth for ablation studies.

% \subsubsection{Transfer learning for INSPIRE dataset}
% \label{sec:transfer-INSPIRE}

\vspace{0.5em}
\noindent\textbf{Transfer learning for the INSPIRE dataset:} As we noted above, the INSPIRE dataset does not have ground-truth vessel segmentation and AV masks available \cite{inspire_dataset-tang2011robust}. We only had access to a labeled sparse graph specifying the topology of the vessels in these images \cite{inspire_av_dataset}. Thus, we trained a U-Net model on similar datasets---specifically, DRIVE \cite{drive_dataset}, HRF \cite{HRF_dataset} and CHASEDB \cite{chasedb_dataset})---with available ground-truth data---either segmentation, AV labels, or both in the case of DRIVE. We then applied this network to INSPIRE, as a form of transfer learning. To obtain an AV likelihood map, we first binarized the segmentation map and then assigned an AV label to each segmented pixel based on the nearest edge in the sparse AV-labeled graph. We treated this AV labeling as the AV ground-truth in our experiments.

\subsection{Results}
Tables~\ref{tab:results-drive}-\ref{tab:results-inspire} show our AV classification results for each of the three datasets. Each pair of row represents one of the three pipelines discussed above (ablated or full). These correspond to different combinations of vessel and artery/vein priors, as illustrated in the flowchart in Fig.~\ref{fig-flowchart}. In more detail, SEG refers to the pixel-level segmentation and AV is the artery-vein likelihood map. The different superscripts and subscripts indicate different versions of the pipeline, namely our full pipeline and the two variants used for ablation studies. The \textit{pmap} subscript indicates that a variant used the likelihood map output produced by a neural network, while \textit{gt} means that a variant used the ground truth data for segmentation or AV labels, as we explained above. Finally, the $0$ superscript corresponds to the AV labels before graph optimization (i.e., just based on the initial, pixel-level labels), while $*$ shows the results after extracting the graph and using it to update the labels. For example, $SEG_{gt} + AV_{pmap}$ means we used the ground-truth segmentation and the AV priors generated by our U-Net network. 

% \textbf{combination} column in tables \ref{tab:results-inspire_same}, \ref{tab:results-inspire} we do not have an entry for $SEG_{gt}$. We then binarize the likelihood map, and for each vessel pixel in this binary image, we assign AV label based on nearest-neighbor strategy with this sparse A/V label topology graph. We treat this AV labeling as AV-segmentation ground-truth $AV_{gt}$. Thanks to the library \cite{easytorch} for facilitating seamless training and evaluation for transfer learning tasks.

% \subsection{Topology Extraction and AV label propagation}

For our main analysis, we calculated three different precision, recall, and F1 scores, listed in different columns groups with superscripts|$g$,  $iseg$, and $seg$. Intuitively, each group measures performance at either the graph or the pixel level. In more detail, columns marked as $g$ correspond to a \textit{node-level} comparison. In other words, we check which percentage of the nodes in the graph have the correct label, both based on the prior (the rows marked with a $0$ superscript) and after label propagation (the rows marked with the $*$ superscript).

% , consisting of the $F_1$ score of the nodes of graph $g$

% \ref{subsec-method-top_extraction} labeled based on corresponding prior in \textbf{combination} and $g$ mapped with ground-truth. As such, we label each node of $g$ with either \textit{Artery} or \textit{Vein}. As mentioned before, the combination in bold is after we apply our AV propagation algorithm. 

% \subsubsection{Common pixels comparison}
% \label{sec:iseg_INSPIRE}
The $iseg$ columns measure the precision, recall, and F1 score only on the pixels that are shared by the vessel segmentation and the ground truth segmentation. In other words, our U-net segmentation network has both false positives (pixels marked as vessel that are background) and false negatives (pixels marked as background that are vessel). The former have no ground truth AV labels because they are not part of a vessel. As such, here we only determine the classification rate given the pixels that are shared by both the U-net likelihood map and the ground truth segmentation (i.e., true positives). In this scenario, the pixels under consideration will only have either an \textit{artery} or \textit{vein} label since the intersection omits the background.

% nd We also assess by AV labeling a binarized vessel likelihood map by using Nearest Neighbor approach as mentioned above. We the take only the common pixels between such labels and ground truth, and calculate $F_1$ score. Here, the pixel under consideration will only have either \textit{Artery} or \textit{Vein} label since taking intersection omits background.

% \subsubsection{Complete AV-label comparison}
Finally, columns marked with a $seg$ superscript consider \textit{all} pixels, both vessel and background. In this case, we have 3 possible labels for each pixel|artery, vein, and background. We then calculated micro-precision, micro-recall, and micro-F1 scores across the three classes. 

% The table lists the average score across the three classes.

% Then we use the manual AV ground-truth to calculate $F_{seg}$ score. We keep record of this score  before and after applying propagation algorithm. 

\begin{table}[ht!]
\begin{center}
    \small
    \caption{Parameter used in all algorithms.}
    \label{tab:exp-parameters}
    \begin{tabular}{|l|c|c|}
    \hline
    \textbf{Parameter} & \textbf{DRIVE/WIDE} & \textbf{IOSTAR}  \\
    \hline
    T & $range[250, 25, -20]$ & $range[250, 25, -20]$ \\
    \hline
    $w_v$ & 5.0 & \textbf{3.0} \\
    \hline
    $w_c$ & 1.0 & 1.0 \\
    \hline
    $w_w$ & 11.0 & \textbf{5.0} \\
    \hline
    $\gls{diam_lim}$ & 10 & 10 \\
    \hline
    $L$ & 11 & \textbf{15} \\
    \hline
    $\gls{dijks2_iters}$ & 3 & 3 \\
    \hline
    $\gls{dist_r}$ & 1.0 & \textbf{2.0} \\
    \hline
    $\gls{dijks2_k_nbrs}$ & 3 & 3 \\
    \hline
    $\gls{c_jump}$ & 3.0 & \textbf{5.0} \\
    \hline
    $\gls{onh_r}$ & 50 & \textbf{60} \\
    \hline
    $\gls{b_th}$ & 2.0 & 2.0 \\
    \hline
    $\gls{a_th}$ & 0.2 & 0.2 \\
    \hline
    $\gls{a_ce}$ & 0.8 & 0.8 \\
    \hline
    $\gls{a_1}$ & 1.0 & 1.0 \\
    \hline
    $\gls{a_2}$ & 1.0 & 1.0 \\
    \hline
    $\gls{a_3}$ & 0.5 & 0.5 \\
    \hline
    $\gls{a_4}$ & 0.5 & 0.5 \\
    \hline
    $\gls{prop_score_thr}$ & 0.75 & 0.75 \\
    \hline
    
    \end{tabular}
\end{center}
    
\end{table}

\begin{table*}[ht!]
\caption{Topology estimation results on DRIVE dataset:}
\label{tab:results-drive}
\setlength{\tabcolsep}{1.5pt}

\begin{normalsize}
\begin{center}

\setlength{\tabcolsep}{0.4em} % for the horizontal padding
\renewcommand{\arraystretch}{1.25}% for the vertical padding
\begin{tabular}{|l|l|ccc|ccc|ccc|}
\hline

&& \multicolumn{3}{c|}{Precision} & \multicolumn{3}{c|}{Recall} & \multicolumn{3}{c|}{F$_1$} \\
\hline

\textbf{Pipeline} & \textbf{Combination} & \textbf{P$^g$} & \textbf{P$^{iseg}$} & \textbf{P$^{seg}$} & \textbf{R$^{g}$} & \textbf{R$^{iseg}$} & \textbf{R$^{seg}$} & \textbf{F$^{g}_1$} & \textbf{F$^{iseg}_1$} & \textbf{F$^{seg}_1$} \\
\hline

U-net seg. \& AV GT & SEG$_{pmap}^{0}$ + AV$_{gt}^{0}$ & 0.9181 & 0.9898 & 0.9708 & 0.8872 & 0.9895 & 0.9708 & 0.8941 & 0.9896 & 0.9708  \\

&\textbf{SEG$_{pmap}^{*}$ + AV$_{gt}^{*}$} & 0.9325 & 0.9638 & 0.9684 & 0.9274 & 0.9643 & 0.9684 & 0.9287 & 0.9638 & 0.9684 \\
\hline

GT seg. \& U-net AV &SEG$_{gt}^{0}$ + AV$_{pmap}^{0}$ & 0.7814 & 0.8738 & 0.9649 & 0.7822 & 0.8723 & 0.9649 & 0.7784 & 0.8710 & 0.9649 \\

&\textbf{SEG$_{gt}^{*}$ + AV$_{pmap}^{*}$} & 0.8171 & 0.8806 & 0.9653 & 0.8180 & 0.8798 & 0.9653 & 0.8143 & 0.8781 & 0.9653  \\
\hline

Full pipeline & SEG$_{pmap}^{0}$ + AV$_{pmap}^{0}$ & 0.7780 & 0.8815 & 0.9625 & 0.7808 & 0.8795 & 0.9625 & 0.7754 & 0.8783 & 0.9625 \\

&\textbf{SEG$_{pmap}^{*}$ + AV$_{pmap}^{*}$} & 0.8140 & 0.9015 & 0.9631 & 0.8174 & 0.9057 & 0.9631 & 0.8126 & 0.9036 & 0.9631 \\
\hline

\end{tabular}
\end{center}
\end{normalsize}
\end{table*}

\begin{table*}[ht!]
\caption{Topology estimation results on WIDE dataset:}
\label{tab:results-wide}
\setlength{\tabcolsep}{1.5pt}

\begin{normalsize}
\begin{center}

\setlength{\tabcolsep}{0.4em} % for the horizontal padding
\renewcommand{\arraystretch}{1.25}% for the vertical padding
\begin{tabular}{|l|l|ccc|ccc|ccc|}
\hline

&& \multicolumn{3}{c|}{Precision} & \multicolumn{3}{c|}{Recall} & \multicolumn{3}{c|}{F$_1$} \\
\hline

\textbf{Pipeline}&\textbf{Combination} & \textbf{P$^g$} & \textbf{P$^{iseg}$} & \textbf{P$^{seg}$} & \textbf{R$^{g}$} & \textbf{R$^{iseg}$} & \textbf{R$^{seg}$} & \textbf{F$^{g}_1$} & \textbf{F$^{iseg}_1$} & \textbf{F$^{seg}_1$} \\
\hline

U-net seg. \& AV GT&SEG$_{pmap}^{0}$ + AV$_{gt}^{0}$ & 0.9117 & 0.9947 & 0.9753 & 0.8816 & 0.9947 & 0.9753 & 0.8858 & 0.9947 & 0.9753  \\

&\textbf{SEG$_{pmap}^{*}$ + AV$_{gt}^{*}$} & 0.9319 & 0.9670 & 0.9726 & 0.9299 & 0.9663 & 0.9726 & 0.9302 & 0.9665 & 0.9726 \\
\hline

GT seg. \& U-net AV&SEG$_{gt}^{0}$ + AV$_{pmap}^{0}$ & 0.8122 & 0.8850 & 0.9721 & 0.8114 & 0.8825 & 0.9721 & 0.8102 & 0.8822 & 0.9721 \\

&\textbf{SEG$_{gt}^{*}$ + AV$_{pmap}^{*}$} & 0.8739 & 0.9127 & 0.9736 & 0.8724 & 0.9111 & 0.9736 & 0.8715 & 0.9107 & 09736.  \\
\hline

Full pipeline &SEG$_{pmap}^{0}$ + AV$_{pmap}^{0}$ & 0.7982 & 0.8955 & 0.9681 & 0.7989 & 0.8931 & 0.9681 & 0.7967 & 0.8927 & 0.9681 \\

&\textbf{SEG$_{pmap}^{*}$ + AV$_{pmap}^{*}$} & 0.8305 & 0.9079 & 0.9685 & 0.8311 & 0.9092 & 0.9681 & 0.8285 & 0.9085 & 0.9685 \\
\hline

\end{tabular}
\end{center}
\end{normalsize}
\end{table*}

\begin{table*}[ht!]
\caption{Topology estimation results on INSPIRE dataset(Same parameters)}
\label{tab:results-inspire_same}
\setlength{\tabcolsep}{1.5pt}

\begin{normalsize}
\begin{center}

\setlength{\tabcolsep}{0.4em} % for the horizontal padding
\renewcommand{\arraystretch}{1.25}% for the vertical padding
\begin{tabular}{|l|l|ccc|ccc|ccc|}
\hline

&& \multicolumn{3}{c|}{Precision} & \multicolumn{3}{c|}{Recall} & \multicolumn{3}{c|}{F$_1$} \\
\hline

\textbf{Pipeline}&\textbf{Combination} & \textbf{P$^g$} & \textbf{P$^{iseg}$} & \textbf{P$^{seg}$} & \textbf{R$^{g}$} & \textbf{R$^{iseg}$} & \textbf{R$^{seg}$} & \textbf{F$^{g}_1$} & \textbf{F$^{iseg}_1$} & \textbf{F$^{seg}_1$} \\
\hline

U-net seg. \& AV GT&SEG$_{pmap}^{0}$ + AV$_{gt}^{0}$ & 0.9412 & 0.9873 & 0.9921 & 0.9369 & 0.9869 & 0.9921 & 0.9370 & 0.9871 & 0.9921  \\

&\textbf{SEG$_{pmap}^{*}$ + AV$_{gt}^{*}$} & 0.9367 & 0.9447 & 0.9890 & 0.9353 & 0.9445 & 0.9890 & 0.9350 & 0.9439 & 0.9890 \\

\hline

Full pipeline&SEG$_{pmap}^{0}$ + AV$_{pmap}^{0}$ & 0.8105 & 0.8696 & 0.9839 & 0.8092 & 0.8695 & 0.9839 & 0.8080 & 0.8677 & 0.9839 \\

&\textbf{SEG$_{pmap}^{*}$ + AV$_{pmap}^{*}$} & 0.8303 & 0.8697 & 0.9837 & 0.8267 & 0.8673 & 0.9837 & 0.8250 & 0.8653 & 0.9837 \\
\hline

\end{tabular}
\end{center}
\end{normalsize}
\end{table*}

\begin{table*}[ht!]
\caption{Topology estimation results on INSPIRE dataset(With some parameters optimized as in table \ref{tab:exp-parameters})}
\label{tab:results-inspire}
\setlength{\tabcolsep}{1.5pt}

\begin{normalsize}
\begin{center}

\setlength{\tabcolsep}{0.4em} % for the horizontal padding
\renewcommand{\arraystretch}{1.25}% for the vertical padding
\begin{tabular}{|l|l|ccc|ccc|ccc|}
\hline

&& \multicolumn{3}{c|}{Precision} & \multicolumn{3}{c|}{Recall} & \multicolumn{3}{c|}{F$_1$} \\
\hline

\textbf{Pipeline} & \textbf{Combination} & \textbf{P$^g$} & \textbf{P$^{iseg}$} & \textbf{P$^{seg}$} & \textbf{R$^{g}$} & \textbf{R$^{iseg}$} & \textbf{R$^{seg}$} & \textbf{F$^{g}_1$} & \textbf{F$^{iseg}_1$} & \textbf{F$^{seg}_1$} \\
\hline

U-net seg. \& AV GT&SEG$_{pmap}^{0}$ + AV$_{gt}^{0}$          & 0.9416 & 0.9865 & 0.9920 & 0.9380 & 0.9863 & 0.9920 & 0.9378 & 0.9864 & 0.9920  \\

&\textbf{SEG$_{pmap}^{*}$ + AV$_{gt}^{*}$} & 0.9450 & 0.9475 & 0.9891 & 0.9437 & 0.9469 & 0.9891 & 0.9435 & 0.9467 & 0.9891 \\

\hline

Full pipeline&SEG$_{pmap}^{0}$ + AV$_{pmap}^{0}$  & 0.8158 & 0.8745 & 0.9841 & 0.8146 & 0.8739 & 0.9841 & 0.8132 & 0.8723 & 0.9841 \\

&\textbf{SEG$_{pmap}^{*}$ + AV$_{pmap}^{*}$} & 0.8427 & 0.8776 & 0.9843 & 0.8391 & 0.8837 & 0.9843 & 0.8379 & 0.8806 & 0.9843 \\
\hline

\end{tabular}
\end{center}
\end{normalsize}
\end{table*}

% We have evaluated the performance of topology estimation using Precision, recall and $F_1$ scores. Using same parameters(tab. \ref{tab:exp-parameters}) 

We used the same parameters, listed in Tbl.~\ref{tab:exp-parameters}, for both DRIVE and WIDE. As our results in these tables show, our graph pipeline yielded significant classification improvements in both datasets. Specifically, we can see almost a 5$\%$ improvement in node-level labels in both DRIVE and WIDE after updating the labels using our graph-based propagation method. Similarly, we can also see a substantial improvements in common-pixel, and full segmentation mask comparisons. This shows that our graph-based topology matches the underlying vasculature well, in that the labels propagated using our estimated topology are more accurate than those obtained by a deep learning method (U-net) alone.

% have indeed captured the notion of a vascular topology. 

Additionally, we can see similar improvements with the INSPIRE dataset, especially after a few parameter adjustments. Such adjustments were necessary because INSPIRE images are larger than DRIVE or WIDE and have different color/contrast characteristics. As mentioned before, we used transfer learning to obtain a vessel likelihood map for this dataset. Despite the lack of ground truth, however, our pipeline was still able to yield significant improvements after label propagation ($\sim$4\% with some parameter optimization as shown in Tbl.~\ref{tab:exp-parameters}). Even in the case where we used the same parameters as in the other two datasets, our pipeline was still able to improve on the U-net priors (see Tbl.~\ref{tab:results-inspire}).

\begin{table*}[ht!]
\caption{A/V segmentation result of existing techniques}
\label{tab:literature-results}
\setlength{\tabcolsep}{1.5pt}
\begin{center}
\begin{small}
    \begin{tabular}{|l|ccc|ccc|ccc|}
        \hline
        & \multicolumn{3}{c|}{DRIVE} & \multicolumn{3}{c|}{WIDE} & \multicolumn{3}{c|}{INSPIRE} \\
        \hline
        
        \textbf{Method} & \textbf{BACC} & \textbf{SEN} & \textbf{SPE}  & \textbf{BACC} & \textbf{SEN} & \textbf{SPE} & \textbf{BACC} & \textbf{SEN} & \textbf{SPE} \\
        \hline 
        
        \cite{graphav1_6517259} & 0.870 & 0.90 & 0.84 & - & - & - & 0.865 & 0.910 & 0.860 \\
        \hline
        
        \cite{7120990} & 0.935 & 0.930 & 0.941 & 0.910 & 0.909 & 0.910 & 0.915 & 0.902 & 0.909 \\
        \hline
        
        \cite{cnn_mst_8309054} & 0.927 & 0.923 & 0.931 & - & - & - & - & - & - \\ 
        \hline
        
        \cite{multi_task_https://doi.org/10.48550/arxiv.2007.09337} & 0.944 & 0.934 & 0.955 & - & - & - & 0.918 & 0.924 & 0.913 \\
        \hline
        
        \cite{Hu2021} & 0.955 & 0.936 & 0.974 & - & - & - & - & - & - \\
        \hline
        
        \textbf{Full Pipeline} & 0.890 & 0.9057 & 0.8758 & 0.902 & 0.9092 & 0.8965 & 0.870 & 0.8837 & 0.8561 \\
        \hline
    \end{tabular}
\end{small}
\end{center}
\end{table*}

Finally, Table~\ref{tab:literature-results} shows a comparison of our fully automatic method to existing, semi-automatic AV segmentation techniques. Here we list the accuracy, sensitivity, and specificity of our method since those are the primary metrics used across the different papers. We list these various results to put our performance in context, but it is important to note that the results across the different papers are not directly comparable. Some papers only list results for centerline pixels while others show results for all the segmented pixels. In addition, the classification accuracy of all these methods depends on how much of the vasculature they identify as being part of a vessel (i.e., vessel recall). Performing AV classification on an undersegmented vasculature, i.e., one where only the main vessels have been segmented, will naturally yield better results than trying to classify both the large and small vessels. All the methods listed in Tbl.~\ref{tab:literature-results} used different recall levels in their vessel segmentation, which further complicates direct comparisons. That being said, we note that our fully automatic approach was able to achieve AV classification results comparable to techniques that are semi-automated and have a much lower vessel recall. We believe that our proposed approach will be able to match these semi-automated techniques given more training data.

% Due to the high skew in the ratio of vessels vs. background pixels,

% \textcolor{red}{Table \ref{tab:literature-results} shows the result for Artery-Vein classification. The primary goal of in this paper being to extract the complete topology of a fundus image, we have shown that having good priors yields better results. As a baseline, we used a vanilla UNet for vessel prior and Artery-Vein priors, and generate topology with that. Next, we have ablated one of the priors with ground truth and achieved significant improvements in Artery-Vein segmentation showing that better segmentation techniques for prior generation will definitely yield better results.}

% with some parameter optimized(Tab. \ref{tab:exp-parameters}). some improvement with the same parameters

% (Tab. \ref{tab:results-inspire_same}), and 

\section{Discussion}
\label{sec:discussion}
Most deep learning and machine learning techniques are trained to only diagnose the absence or presence of a single disease. Different models are used for different diseases because the most informative pathological markers differ between diseases. As such, each diagnostic system requires significant time, effort, and resources to train. Furthermore, since each system is trained in isolation, we cannot compare features across models to gain further insight into a patient's health. 

In this work, we presented a general-purpose topology extraction method for retinal fundus images. Our ultimate goal is to leverage this graph-based representation to effectively extract all relevant vascular features using a single system. This system will, in turn, allow us to diagnose multiple diseases with the same, explainable features. For example, we could use our pipeline to help estimate the artery/vein ratio, tortuosity, bifurcation statistics, distribution of disconnected vessels, etc. Our current work is an important milestone in this direction because we have shown that we can extract the vasculature and artery-vein labels using a single, fully end-to-end system. No manual intervention is needed to correct mistakes in the topology or in the labels. In addition, as our ablation studies show (Tbls.~\ref{tab:results-drive}, \ref{tab:results-wide}, and \ref{tab:results-inspire}), the performance of our system improves given better priors. This is encouraging because it shows that our topology model accurately captures how vessels are distributed in the retina. In other words, our graph editing operations rarely introduce errors that were not present in the initial estimate. Furthermore, we have even shown that our system trained on the DRIVE dataset works well on a second dataset (INSPIRE) with little-to-no parameter calibration, further validating the generality of our graph-based model.

% Overall, we believe that this work will enable a more principled path for automated fundus image diagnosis. 

In addition, our graph representation may help physicians compute novel features of interest. For example, in Fig.~\ref{fig-vonh} we can see how our flow-assignment algorithm detects possible points where vessel are broken (shown as large yellow nodes in the image). These gaps in the graph usually reflect poor lighting conditions, but they could also signal locations of concern in the vasculature itself (e.g., where there might be an obstruction or a hemorrhage). Potentially, the quality of the extracted graph itself may help ophthalmologists gauge the patient's overall health.

% but they could also The size of the nodes in this figure depicts if a natural blood path has been broken or obstructed. For example, in fig. \ref{fig-vonh}, we can see disconnected vessel end point get more visits than the other ones(near ONH center left of first image,  top right of second and third image). We have only used the directed graph obtained from the flow-assignment algorithm, however intuitive depiction blood flow obstruction could be used in future research as a important factor in itself.

\section{Conclusion and Future Work}
\label{sec:conclusion-and-future-work}
Automated disease diagnosis is not fully trusted in the medical profession, in large part because current methods are not explainable enough. Therefore, our goal is to develop fundus analysis techniques that are both diagnostically useful and understandable to a human operator. To that end, in this work we presented a framework for automatically extracting and labeling the entire retinal vasculature given a single fundus image. We believe our graph-based representation of the vasculature will open new avenues for local and global vascular feature analysis. We have also shown that our pipeline generalizes well to novel data, as shown by our transfer learning results on INSPIRE. Also, in addition to extracting the vessels themselves, our method also identifies other features of interest, including the pseudo-ONH shown in Fig. \ref{fig-vonh}. It is also important to note that our pipeline only requires the vessel probability map to generate a directed topology graph. We have shown how this graph can be used for A/V label propagation; however, we believe that one could similarly use our graph for estimating other features of interest, including vessel tortuosity, bifurcation, etc., with minimal-to-no changes. We plan to explore estimating these additional features in future work. These features may, in turn, lead to more robust and explainable detection tools for a number of diseases, including glaucoma, diabetic retinopathy, and macular degeneration. We also intend to explore this form of automated diagnosis in future work.

% Figure~\ref{fig-flowchart} shows additional downstream task that our pipeline may facilitate in the future.

% It also enables analysis of the overall vascular system analysis in fundus image. In fact, a prominent future work is to use the topology graph to detect features like turtuosity, bifurcation and optimal blood flow in early detection of diseases like Diabetic Retinopathy. 

%%Harvard
% \bibliographystyle{model2-names.bst}\biboptions{authoryear}
\bibliographystyle{plain}
\bibliography{refs}
\end{document}